\documentclass[a4paper,10pt]{article}
\newcommand{\EE}{\mathbb{E}}
\newcommand{\widebar}[1]{\overline{#1}}

\newcommand{\drm}{\, \text{\rmfamily d}}
\newcommand{\Vpen}{V^\epsilon}
\newcommand{\Wpen}{W^\epsilon}
\newcommand{\wpen}{w^\epsilon}
\newcommand{\upen}{u^\epsilon}
\newcommand{\xpen}{x^\epsilon}
\newcommand{\rpen}{r^\epsilon}
\newcommand{\taupen}{\tau^\epsilon}
\newcommand{\stoptime}{\tau^\epsilon \wedge t}
\newcommand{\sqrteps}{\epsilon^{1/2}}
\newcommand{\lar}{L\!\,a\!\,r\!g\!\,e}

\newcommand{\Vl}{V}
\newcommand{\Hl}{H}
\newcommand{\Bopl}{B}
\newcommand{\al}{a}
\newcommand{\bl}{b}
\newcommand{\Ol}{\Omega}
\newcommand{\brackl}{)}

\newcommand{\constgam}{\Gamma^*\!(t)}
\newcommand{\xst}{x^*\!(t)}
\newcommand{\Sst}{{S}^*\!(t)}

\newcommand{\beq}{\begin{equation}}
\newcommand{\eeq}{\end{equation}}

\newcommand{\bs}{\begin{slide}{}}
\newcommand{\es}{\end{slide}}

\newcommand{\bc}{\begin{center}}
\newcommand{\ec}{\end{center}}

\newcommand\cyr{%
  \renewcommand\rmdefault{wncyr}%
  \renewcommand\sfdefault{wncyss}%
  \renewcommand\encodingdefault{OT2}%
  \normalfont
  \selectfont}
\DeclareTextFontCommand{\textcyr}{\cyr}

\newcommand\eq[1] {(\ref{#1})}

\newtheorem{definition}{Definition}[section]

\newcommand{\beqa}{\begin{eqnarray}}
\newcommand{\eeqa}{\begin{eqnarray}}

\newcommand\barint{- \kern-1.0495em \int}
\newcommand{\romd}{\mathrm{d}}
\newcommand{\rome}{\mathrm{e}}

\newcommand{\Ga}{\alpha}

\newcommand{\Gs}{\sigma}

\newcommand{\overliner}{\begin{eqnarray}}
\newcommand{\earr}{\end{eqnarray}}
\newcommand{\overlinern}{\begin{eqnarray*}}
\newcommand{\earrn}{\end{eqnarray*}}
\newcommand{\prt}{\partial}

\def\pda#1#2{{\frac{\partial {#1}}{\partial {#2}}}}
\def\pdb#1#2{{\frac{\partial^2{#1}}{ \partial {#2}^2}}}

\def\cpda#1#2{{\partial {#1}/ \partial {#2}}}
\def\cpdb#1#2{{\partial^2{#1}/ \partial {#2}^2}}

\def\pda#1#2{{\frac{\partial{#1}}{\partial{#2}}}}
\def\cpda#1#2{\ifinner
   {\partial{#1}/\partial{#2}}
\else
   {\frac{\partial{#1}}\frac{\partial{#2}}}
\fi}

\def\cpdb#1#2{\ifinner
   {\partial^2{#1}/\partial{#2}^2}
\else
   {\partial^2{#1}\over\partial{#2}^2}
\fi}

\def\cpdc#1#2#3{\ifinner
    {\partial^{#3}{#1}/\partial{#2}^{#3}}
\else
    {\partial^{#3}{#1}\over\partial{#2}^{#3}}
\fi}

\def\cpdm#1#2#3{\ifinner
    {\partial^2{#1}/\partial{#2}\partial{#3}}
\else
   {\partial^2{#1}\over\partial{#2}\partial{#3}}
\fi}

\def\ctda#1#2{\ifinner
   {\hbox{d}{#1}/\hbox{d}{#2}}
\else
   {\hbox{d}{#1}\over\hbox{d}{#2}}
\fi}

\def\ctdb#1#2{\ifinner
   {\hbox{d}^2{#1}/\hbox{d}{#2}^2}
\else
   {\hbox{d}^2{#1}\over\hbox{d}{#2}^2}
\fi}

\def\ctdc#1#2#3{\ifinner
    {\hbox{d}^{#3}{#1}/\hbox{d}{#2}^{#3}}
\else
    {\hbox{d}^{#3}{#1}\over\hbox{d}{#2}^{#3}}
\fi}

\def\cmda#1#2{\ifinner
   {\hbox{D}{#1}/\hbox{D}{#2}}
\else
   {\hbox{D}{#1}\over\hbox{D}{#2}}
\fi}

\def\doubleint{\mathop{\int\kern -0.8em\int}\nolimits}

\def\harf{
\ifinner
    {\frac{1}{2}}
\else
    {\hbox{$\frac{1}{2}$}}
\fi
}

\def\quater{
\ifinner
    {\frac{1}{4}}
\else
    {\hbox{$\frac{1}{4}$}}
\fi
}
\def\sixth{
\ifinner
    {\frac{1}{6}}
\else
    {\hbox{$\frac{1}{6}$}}
\fi
}
\def\twforth{
\ifinner
    {\frac{1}{24}}
\else
    {\hbox{$\frac{1}{24}$}}
\fi
}

\usepackage{a4wide}
\usepackage{amssymb,epsfig,amsmath}
\usepackage{graphicx,subfig,color}
\graphicspath{{./Figs/}}
\usepackage{latexsym, relsize}

\title{The Effect of Non-Smooth Payoffs on the Penalty Approximation of American Options}

\author{S. D. Howison\thanks{Mathematical Institute and Oxford-Man Institute of Quantitative Finance, University of Oxford, OX1 3LB, Oxford, UK ({\tt [\,howison\,,\,reisinge\,,\,witte\,]\,@\,maths.ox.ac.uk})} \and
C. Reisinger\footnotemark[1] \and
{J. H. Witte\footnotemark[1]} \thanks{J.~H.~Witte acknowledges support from Balliol College and the Oxford-Man Institute, University of Oxford, and the UK Engineering and Physical Sciences Research Council (EPSRC)}
}

\begin{document}
\maketitle
\newcommand{\slugmaster}{%
\slugger{siads}{xxxx}{xx}{x}{x--x}}

\newtheorem{theorem}{Theorem}[section]
\newtheorem{la}[theorem]{Lemma}
\newtheorem{cor}[theorem]{Corollary}
\newtheorem{remark}[theorem]{Remark}
\newtheorem{prob}[theorem]{Problem}	
\begin{abstract}
This article combines various methods of analysis to draw a comprehensive picture
of penalty approximations to the value, hedge ratio, and optimal exercise strategy of American options.
We use matched asymptotic expansions to characterise the boundary layers between exercise
and hold regions, and to compute first order corrections for representative payoffs
on a single asset following a diffusion or jump-diffusion model.
Furthermore, we demonstrate how the viscosity theory framework in \cite{jakobsen06} can be applied 
to derive upper and lower bounds on the option value. This analysis confirms the higher order of accuracy
in the penalty parameter for convex payoffs (compared to the general case) seen earlier in numerical tests and from asymptotic expansions.
In a small extension to \cite{Lions_ApplVarIneqStochControl}, we derive weak convergence rates also for option sensitivities for convex payoffs under
jump-diffusion models.
Finally, we outline applications of the results, including accuracy improvements by extrapolation.
\end{abstract}


\hspace{-.45cm}\textit{Key Words:} American Option, Jump-Diffusion Model, Penalty Method,
Penalization Error, Non-Smooth Payoff\bigskip

\hspace{-.45cm}\textit{2010 Mathematics Subject Classification:} 60G40, 47G20\bigskip

\section{Introduction}\label{Section_Introduction}

An American option is a financial instrument that gives its holder the right to claim a specified payoff on an asset at any time up to a certain date.
Pricing an American option involves determining an optimal exercise strategy in addition to the price itself.
For simplicity, we discuss first the Black-Scholes setting (cf.\,\cite{BlackScholesOrigPaper}), i.e., where the stock price follows
\begin{equation}
\text{d}S_t/S_t = \mu\,\text{d}t + \sigma \, \text{d}W_t,
\end{equation}
where $\sigma$ is the volatility, $\mu$ the drift rate, and $W$ a standard Brownian motion.\bigskip

\bigskip

There are two main equivalent formulations of this problem:
a probabilistic one based on optimal stopping, and a deterministic one
in the form of a linear complementarity problem (free boundary problem).
The optimal stopping formulation was first introduced
in \cite{Bensoussan_TheoryOptionPricing} and \cite{Karatzas_AmericanOptions};
a concise outline can be found in \cite{Shreve_ContTimeModels}.

\bigskip

In \cite{OptionPricing}, it is described how an American option can be priced using a linear 
complementarity problem (LCP)
\begin{eqnarray}
\label{lcp}
\min(-\mathcal{L}_\mathrm{BS} V, V-\Psi) = 0,
\end{eqnarray}
where $\Psi$ is the payoff and
$\mathcal{L}$ is the Black-Scholes operator
\begin{equation}
  \mathcal{L}_\mathrm{BS} V
:= \pda{V}{t} + \harf\Gs^2S^2\pdb{V}{S}
+(r-q) S\pda{V}{S}-rV,
\end{equation}
where $r$ is the risk-free interest rate and $q$ a continuously paid dividend yield. 
The relation between optimal
stopping times and PDEs is further analysed in \cite{JailletLamberton_VariationalInequalities}.
\bigskip

In this paper, we are concerned
in particular with the effects of 
the payoff function on a 
so-called
\emph{penalty approximation}
to the value of such an option.
Penalty approximations are useful both for the analysis \cite{Lions_ApplVarIneqStochControl, Zhang_AmericanJumpPaper}
and numerical analysis \cite{jakobsen06} of the limiting problem, but
also lend themselves to arguably the most efficient numerical approximation methods presently available
for American option valuation \cite{ForsythQuadraticConvergence}.
\bigskip

Penalisation of (parabolic) variational inequalities is classical (cf.\,\cite{Lions_ApplVarIneqStochControl}).
The canonical penalty approximation of (\ref{lcp}) is
\begin{eqnarray}
\label{pen}
-\mathcal{L}_\mathrm{BS} \Vpen = \frac{1}{\epsilon} \max(\Psi-\Vpen,0)
\end{eqnarray}
for $\epsilon>0$  (cf.\,\cite{Bensoussan_TheoryOptionPricing}).
The penalty term on the right-hand side is only active when $\Vpen<\Psi$, and then it
serves to push $\Vpen$ upwards towards the payoff.\bigskip

In the context of American options, in chronological order, \cite{Wang_PowerPenalty_LCP_AmericanOption,Zhang_ConvergenceAnalysis_MonotonicPenalty,ZhangWang_PowerPenalty_TwoAssetAmerican} study the penalisation error for the Black-Scholes model and different penalty terms, and \cite{Achdou_AnInverseProblem_VolatilityAmericans} uses penalisation implicitly to solve a calibration problem;
\cite{Forsyth_PenaltyStochVol,ForsythQuadraticConvergence,Nielsen_PenaltyAndFrintFixing_AmericanOptions,Forsyth_JumpPenalty,Nielsen_PenaltyAmericanMultiAsset} introduce penalty approximations as a means of solving the discretised variational inequality.

\bigskip

We first address the question of what a relevant measure of accuracy should be. This clearly depends on
what the solution will be used for.

\subsection*{Effect of the Penalisation Error on Pricing and Hedging}

Hedging American options requires knowledge of the hedge ratio,
i.e., the amount of stocks held short in the hedging portfolio per unit long position in the option before it is exercised.
In the complete market case of the Black-Scholes model,
the hedge ratio is the so-called \emph{Delta}, 
$
\Delta_t = ({\partial V}\!\! \,/{\partial S}) (S_t,t).
$\bigskip

Because we do not know the exact option value, but only its penalty approximation, we are exposed to three sources of error if we, say, buy
and hedge an American option:

\bigskip
\begin{itemize}
 \item[a)] We bid the -- lower, as we shall see -- price $\Vpen(S_0,0)$ instead of $V(S_0,0)$ for the option at the outset.
 \item[b)] We hedge with the wrong hedge ratio 
\[
\Delta_t^\epsilon = \pda{\Vpen}{S}(S_t,t)
\]
instead of the exact Delta $\Delta_t$.
\item[c)] We exercise at the wrong time
\[
\taupen := \inf\{0\le t\le T: \Vpen(S_t,t)\le \Psi(S_t)\},
\]
which is no later than the optimal exercise time $\tau$, since $\Vpen(S_t,t) \le V(S_t,t)$.
\end{itemize}

\bigskip
The values $X_t$ and $X_t^{\epsilon}$ of the corresponding hedge portfolios at $t<\taupen<\tau$ are
\begin{eqnarray*}
X_t &=& X_0 + \sigma \int_0^t \Delta_u S_u {\rm e}^{r (t-u)} \, \text{d}W_u^{Q}, \\
X_t^{\epsilon} &=& X_0^{\epsilon} + \sigma \int_0^t \Delta_u^{\epsilon} S_u {\rm e}^{r (t-u)} \, \text{d}W_u^{Q},
\end{eqnarray*}
by a classical replication argument (e.g., \cite{Shreve_ContTimeModels}),
where $\text{d}W_t^{Q} = \text{d}W_t + (\mu-r)/\sigma \;\! \text{d}t$ is the increment of a standard Brownian motion under the risk-neutral measure $Q$.
\bigskip

Consider the stopping time $\stoptime$. 
The stochastic integrals above are semi-martingales, and over a fixed finite time interval true martingales.
Then,
by the Optional Stopping Theorem,
\[
\mathbb{E}^Q[X_{\stoptime}^{\epsilon}-X_{\stoptime}] = X_0^{\epsilon}-X_0
\]
and, by It{\^o isometry},
\begin{eqnarray}
\nonumber
\mathbb{V}^Q[X_{\stoptime}^{\epsilon}-X_{\stoptime}] &=& \mathbb{E}^Q \int_0^{\stoptime} \sigma^2 S_u^2 {\rm e}^{2 r (t-u)} (\Delta_u^{\epsilon}-\Delta_u)^2 \, \text{d}u \\
\nonumber
&\le& \mathbb{E}^Q \int_0^T \sigma^2 S_u^2 {\rm e}^{2 r (t-u)} (\Delta_u^{\epsilon}-\Delta_u)^2 \, \text{d}u \\
\nonumber
&=& \int_0^T \sigma^2 {\rm e}^{2 r (t-u)} \mathbb{E}^Q [S_u^2 (\Delta_u^{\epsilon}-\Delta_u)^2] \, \text{d}u \\
\nonumber
&\le& \sigma^2  {\rm e}^{2 r T} \int_0^T \int_0^\infty  p(S_0,0;S,u) \, S^2 \left(\pda{\Vpen}{S}(S,u)-\pda{V}{S}(S,u) \right)^2 \, \text{d}S \, \text{d}u \\
\label{semi-norm}
&\le& C \int_0^T \int_0^\infty S \left(\pda{\Vpen}{S}(S,u)-\pda{V}{S}(S,u) \right)^2 \, \text{d}S \, \text{d}u,
\end{eqnarray}
where $p(S_0,0;S,u)$ is the transition density of $(S_t)_{0\le t\le T}$, under $Q$, from $S_0$ at time 0 to $S$ at time $u$.
The last inequality follows because $p \,S$ is bounded.
The variance of the replication error at any time prior to exercise is therefore controlled by a weighted semi-norm given in
(\ref{semi-norm}),
which is related to the $H^1$ norm and is one of the error measures we will consider. (The split of $S^2 p$ in the above step into factors $S$ and $S p$ is somewhat arbitrary at this point and will be useful later.)
\bigskip

Next, the loss incurred by exercising too early is
\begin{equation}
V(S_{\taupen},{\taupen})-\Vpen(S_{\taupen},{\taupen}) = V(S_{\taupen},{\taupen})-
\Psi(S_{\taupen}),
\end{equation}
which is the (positive) difference between true and penalised solution at the (sub-optimal) penalty exercise boundary.
Viewed differently, by hedging with $\Delta^\epsilon$ we are replicating an option which is exercised not at the optimal exercise time, but at the crossing time of an approximate exercise boundary. An alternative measure of error is therefore the maximum distance
\[
\|\Vpen - V\|_\infty = \sup_{0\le t\le T, 0\le S} |V(S,t)-\Vpen(S,t)|,
\]
which is also an upper bound for $|X_0- X_0^{\epsilon}| = V(S_0,0) - \Vpen(S_0,0)$.
We will study the convergence in this norm also.


\subsection*{Extension to Jump Models}

A model which allows the asset price process to jump to reflect the possibility of sudden changes in the market was first recorded in \cite{MertonJumpModel}; an extensive overview and detailed discussion of jump models and their use in modern mathematical finance can be found in \cite{JumpProcesses_RamaCont}.
The pricing of American options in the presence of jumps has been developed and studied
in \cite{Zhang_AmericanJumpPaper,Pham_OptStop_JumpDiff},
which remain the main references on the topic.\bigskip

We consider models where 
the underlying asset
follows a jump-diffusion process,
\begin{equation}
\label{jump-diff}
\text{d}S_t/S_t = \mu \, \text{d}t + \sigma \, \text{d}W_t + (J-1) \, \text{d}N_t,
\end{equation}
where
$J$ is a random jump amplitude with values in $[0,\infty)$, and $N$ a compound Poisson process with jump rate $\lambda\ge 0$.
The special case $\lambda=0$ recovers the Black-Scholes model.\bigskip

Under the assumption that jump risk is unpriced, the value of an American option under jump diffusion can still be described by an equation of the type (\ref{lcp}), but with
\begin{equation}
  \mathcal{L}_\mathrm{BSJ} V
:= \pda{V}{t} + \harf\Gs^2S^2\pdb{V}{S}
+(r-q-\omega \lambda) S\pda{V}{S}-rV
+ \lambda\EE[V(JS,t)-V(S,t)],
\label{BSJOP}
\end{equation}
where the expectation is taken with respect to the jump size $J$, for fixed $S$, and
$\omega = \EE[J-1]$.
This can be re-written as a
partial integro-differential equation (PIDE)
in terms of the probability density function $g$ of $J$ via
\[
\EE[V(JS,t)-V(S,t)] = \int_{0}^{\infty} V(SJ,t) g(J) \, \text{d}J \, - \, V(S,t).
\]
It will also be useful to consider the resulting PIDE in $\log$-coordinates, $x=\log(S/S_0)$,
$u(x,t)=V(S_0 \exp(x),t)$,
where
\begin{equation}
\label{jumplogoperator}
\mathcal{L} u = \pda{u}{t} + \harf\Gs^2\pdb{u}{x}
+(r-q-\omega \lambda-\Gs^2/2) \pda{u}{x}-r u
+ \lambda \left[\int_{-\infty}^{\infty} u(x+z,t) \nu(z) \, \text{d}z \, - \, u(x,t)\right]
\end{equation}
in (\ref{lcp}) and $\phi(z) = \Phi(S_0 \exp(z))$ is the new payoff and
$\nu$ the density of $Z = \log(J)$.
The pricing equation is then still (\ref{lcp}), the penalised equation (\ref{pen}), where the operator $\mathcal{L}$ from (\ref{jumplogoperator})
replaces $\mathcal{L}_{BS}$.
\bigskip

We will see that the inclusion of finite activity jumps does not alter the properties of penalty approximations qualitatively, and all the general results later on in
the paper are derived for this class of models.
Some of the specific numerical examples and asymptotic expansions use the Black-Scholes model for ease of exposition.
It will be stated clearly at the start of all sections where this is the case.
\bigskip

\subsection*{Main Findings and Structure of this Paper}

The main contribution of this paper is two-fold: to derive a precise description of the local structure of the penalisation error for relevant example payoffs,
and to give a rigorous analysis of the magnitude of the penalisation error in relevant measures for general payoff classes.

\bigskip

The local structure of the error will be analysed by matched asymptotic expansions and is not visible from the more global functional analysis.
To our knowledge, this is the first study of this behaviour. The (heuristically computed) leading order correction terms will be seen to be in excellent agreement with numerical
computations of the penalisation error, and can thus be used as the basis for accurate extrapolation schemes.

\bigskip

While convergence of the penalised solution for sufficiently smooth obstacles is well established in the 
literature, see, e.g., \cite{Lions_ApplVarIneqStochControl, Zhang_AmericanJumpPaper},
sharp rates of convergence and particularly the effect of gradient discontinuities (i.e., the omni-present `kinks' in option payoffs)
on this rate have not been fully analysed so far.
This becomes important not least when using penalisation as part of a numerical technique for solving the obstacle problem.
The general results here can be classified into two settings: that of
convex kinks between otherwise smooth (usually linear) payoffs, and that of concave kinks.
As concave kinks result in lower convergence order, it is clear that in situations with mixed convexity this behaviour is dominant.

\bigskip

Table \ref{Tab:PenErrRatesSumm} 
provides a summary of the findings in this paper.
It confirms that the convergence order predicted by asymptotic 
analysis is in line with the numerically estimated one in all situations,
while the higher level functional analytic estimates are not always sharp.
\begin{table}[t]
\begin{tabular}{|c||c|c|c|c||c|c|c|c|}
\multicolumn{1}{c}{}
& \multicolumn{4}{c}{Convex kinks} &\multicolumn{4}{c}{Concave kinks}  \\
\hline
 & $L_\infty$ & $W_\infty^1$ & $L_2$ & $H^1$ 
& $L_\infty$ & $W_\infty^1$ & $L_2$ & $H^1$
\\
\hline\hline
Numerical estimate & 1 & 0.55 & 1 & 0.76 & 0.5 & 0.07 & 0.61 & 0.29 \\
Asymptotic expansions & 1 & 0.5 & 1 & 0.75 & 0.5 & $\star$ & 0.5 & 0.25 \\
Functional analysis & 1 & --- & 0.5 & 0.5 & 0.5
& --- & 0 & 0  \\
\hline
\end{tabular}
\caption{Order of convergence in the penalty parameter, for different measures and payoff 
types, and as predicted by different methods of analysis. 
0 indicates convergence, but of no positive order; `---' indicates no known result;
`$\star$' indicates no convergence.
The numerical estimate was obtained by regression of the errors for different numerically computed
penalised solutions. These and the results under asymptotic expansions were computed for representative payoffs
(put and butterfly).
}
\label{Tab:PenErrRatesSumm}
\end{table}

\bigskip

The remainder of this article is organised as follows.
In Section 2, we discuss a few representative examples of typical payoffs and present numerical results as motivation for the following analysis.
In Section 3, we derive the leading order corrections to the penalty solution and the exercise boundary by matched asymptotic expansions,
for the American put and butterfly.
Section 4 generalises the convergence order of the penalisation error of the value to more general classes of convex (order $\epsilon$) and
non-convex (order $\sqrteps$) payoffs, and gives sharp upper and lower bounds on the solution, following the framework of
\cite{jakobsen06} and extending it to jump processes.
Section 5 derives $H^1$ errors, showing that the rate $\sqrteps$ derived in \cite{Lions_ApplVarIneqStochControl} also holds under jump-diffusions and for non-smooth but convex obstacles.
Finally, in Section 6, we discuss the results and their applications; in particular, we show how extrapolation can be used for accuracy improvement.

\section{Different Payoffs and Their Implications}

Part of the appeal of penalty methods as a computational tool is that
the resulting algorithms do not depend on the shape of the payoff. 
In contrast to the formulation as free boundary problem 
(e.g., `front-fixing' methods),
the topology of exercise and continuation regions is irrelevant for the definition
of the penalty approximation and (iterative) solution algorithms based on it.
\bigskip

We will now illustrate how the shape and regularity of the payoff does, however, influence the approximation error.\bigskip

\subsection*{Example Payoffs and their Exercise Strategies}

\begin{figure}[t]
\centering
\includegraphics[scale=.7]{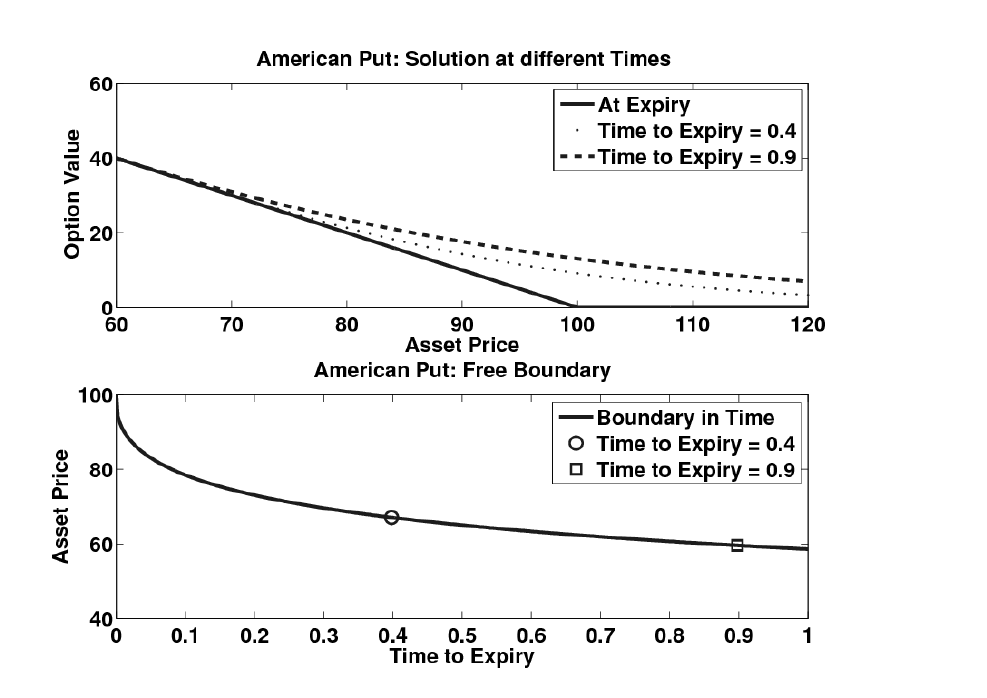}
\caption{The value of an American put at different points in time (above) and the evolution of the corresponding exercise boundary (below).
}
\label{fig:Put}
\end{figure}

\begin{figure}[t]
\centering
\includegraphics[scale=.7]{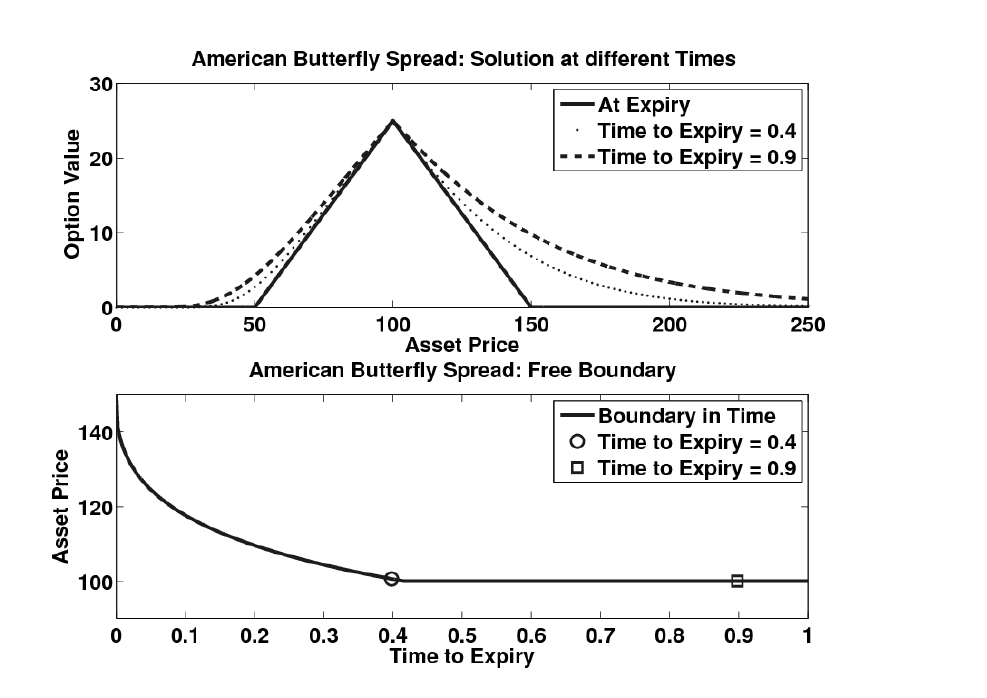}
\caption{The value of an American butterfly spread at different points in time (above) and the evolution of the corresponding exercise boundary (below). Possibly hard to see in the plot, there is (only) one non-trivial free boundary which lies between 100 and 150.
}
\label{fig:Butterfly}
\end{figure}

\begin{figure}[t]
\centering
\includegraphics[scale=.7]{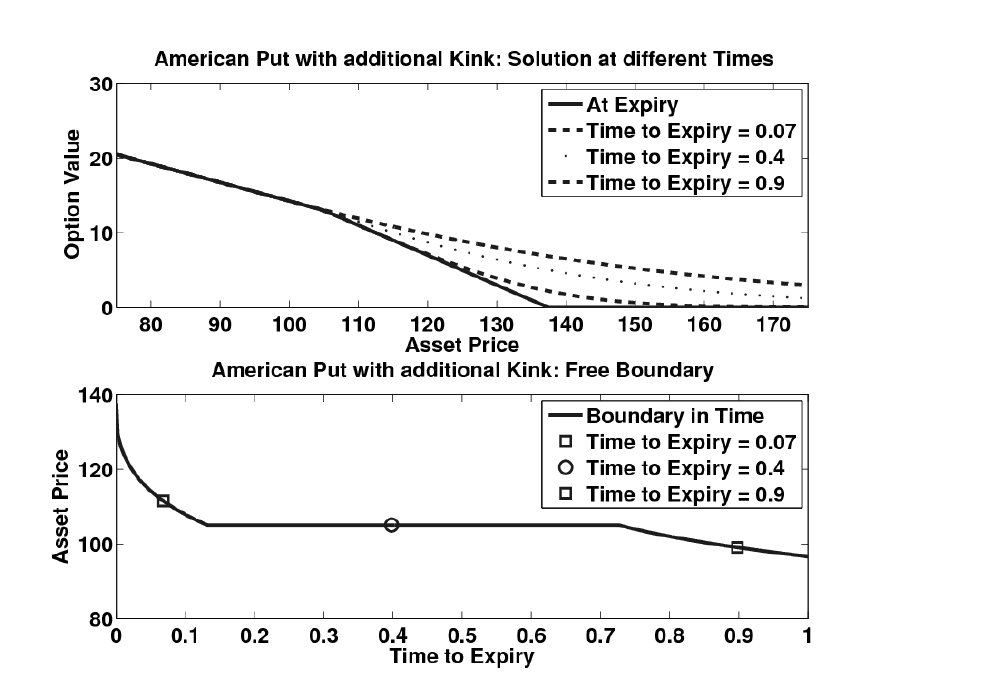}
\caption{
The value function of the `modified' put (with payoff defined in (\ref{modified-put}))
at different points in time (above) and the evolution of the corresponding exercise boundary (below).
}
\label{fig:FunnyPut}
\end{figure}

Two typical payoffs are the standard put payoff (see Fig.~\ref{fig:Put})
\begin{equation}
\Psi(S) = \max(K-S,0),
\end{equation}
with strike $K>0$, and a butterfly spread (see Fig.~\ref{fig:Butterfly})
\begin{equation}
\Psi(S) = 
\max(V_0-\alpha |S-K|,0),
\end{equation}
for some $\alpha, V_0>0$.
We also consider an academic example of a `modified' put (see Fig.~\ref{fig:FunnyPut}), whose piecewise linear payoff
\begin{equation}
\label{modified-put}
\Psi(S) = \alpha \, (\max(K - S,0) - \alpha_1 \max(K_1-S,0))
\end{equation}
is the difference of two put payoffs with strikes $0<K_1<K$, $\alpha>0$ and $0<\alpha_1<1$
(or a sum of a put and a butterfly spread). \bigskip

For American options, typically, an exercise boundary determines the asset price(s) at which 
(for fixed time), the optimal policy switches from holding the option to
exercising.\bigskip

Figures \ref{fig:Put}, \ref{fig:Butterfly}, and \ref{fig:FunnyPut} show value functions with their 
exercise boundaries for different payoffs.
For illustrative purposes, we use a Black-Scholes framework with no dividends, interest rate 
$r=0.05$, volatility $\sigma = 0.4$, and maturity $T=1$.
We will see later that jumps do not change the results qualitatively.\bigskip

We discuss the three examples in turn.\bigskip

\begin{enumerate}
\item
\label{item:standard}
For the standard put, Figure \ref{fig:Put}, a smooth free boundary $S^*(t)$ separates an exercise region $S<S^*(t)$ from a 
hold region $S>S^*(t)$, and decreases strictly as we move away from expiry.
Denote this American put value by $P(S,t)$. Note that, at $S=S^*(t)$,
\beq
\label{FB1}
P(S^*(t),t)=K-S^*(t),\quad \pda{P}{S}(S^*(t),t) =-1, \quad \lim_{S\downarrow S^*(t)} \pdb{P}{S}(S,t) =
\frac{2rK}{\Gs^2{S^*(t)}^2},
\eeq
see, e.g., \cite{OptionPricing}.
Before expiry, the solution is continuously differentiable with a jump in the second
derivative (the \emph{Gamma}) at the exercise boundary.
\item
For the butterfly spread, Figure \ref{fig:Butterfly}, short before expiry, the option value is greater than the payoff
on the call-like side $S<K$, and is similar to the situation in (\ref{item:standard}) above on the put-like side $S>K$, with an exercise boundary between 100 and 150.
The solution smooths the convex kinks which 
the payoff has at 50 and 100; however, for all times, the solution has a kink at 100.
As we move away from expiry, the free boundary on the put-like side at first decreases strictly from 150 and then, having reached $S=100$, remains there.
(One can easily construct a scenario where the free boundary does not reach the location of the concave kink but converges to the exercise boundary of a certain perpetual put.)
\item
For the modified put, Figure \ref{fig:FunnyPut}, there is again a single exercise boundary which separates an exercise region for small $S$ from a hold region.
Before expiry, the solution smooths the convex kink which the payoff has at about 137, but, near expiry, it still has a kink at 105; however, far away from expiry, the solution appears to attach smoothly
to a point on the payoff where $S<105$. As we move away from expiry, the exercise boundary decreases strictly at first, stagnates 
-- until smooth pasting is reached -- and then decreases strictly again. A discussion on this `waiting time' phenomenon in the context of diffusion problems can be found in
\cite{fasanoetal}, to which the present case adds a further example.
\end{enumerate}
\bigskip

\subsection*{Numerical Penalisation Error}
 
We now analyse numerically the penalisation error for these three examples. 
We use a Crank-Nicolson finite difference scheme to discretise (\ref{lcp}) and (\ref{pen}), respectively,
and solve the resulting non-linear discrete system by 
\emph{projected successive over-relaxation} (short PSOR, cf.\,\cite{CryerSystematicOverrelaxation})
in the case of (\ref{lcp}) and a semi-smooth Newton iteration (cf.\,\cite{ForsythQuadraticConvergence})
in the case of (\ref{pen}).

\bigskip

\begin{figure}[t]
\centering
\includegraphics[scale=.4]{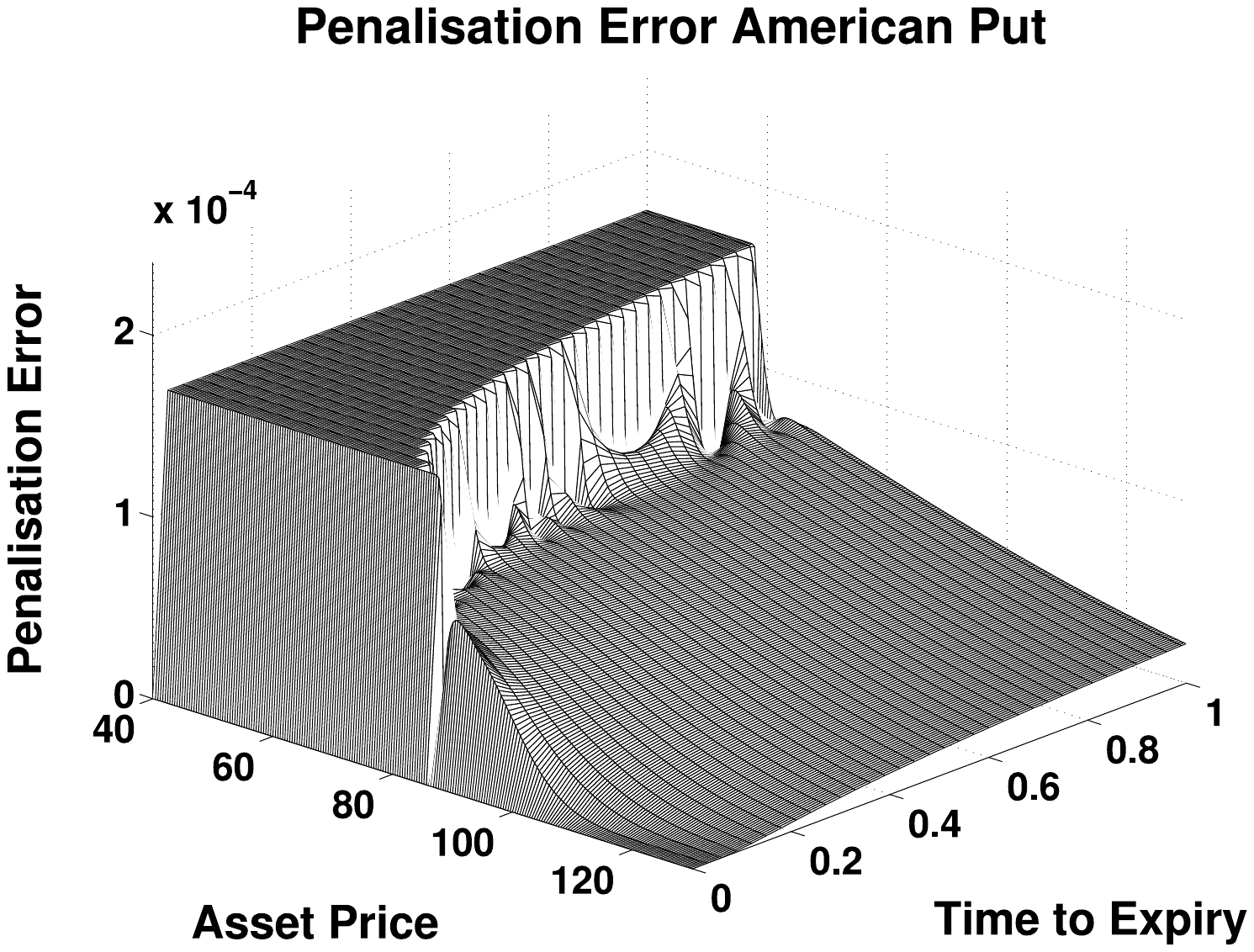}
\caption{Local structure of the penalisation error for the put with parameters $\sigma=0.4$, $r=0.05$, and $\epsilon^{-1}=3\cdot 10^4$. The error is largest, and roughly constant, in the exercise region, and decays rapidly over a small layer around the exercise boundary.}
\label{fig:surfaceput}
\end{figure}

Figure \ref{fig:surfaceput} shows the numerically computed penalisation error for the standard put as a function of $S$ and $t$.
It appears that the error is constant in the exercise region, jumps to about half this value across the exercise boundary, and decays for large $S$.
The irregular behaviour of the plotted error surface close to the exercise boundary is a discretisation artefact due to the movement of the exercise boundary relative to its closest mesh points between subsequent time steps
(see also next paragraph).

\bigskip

\begin{figure}[t]
\centering
\includegraphics[scale=.4]{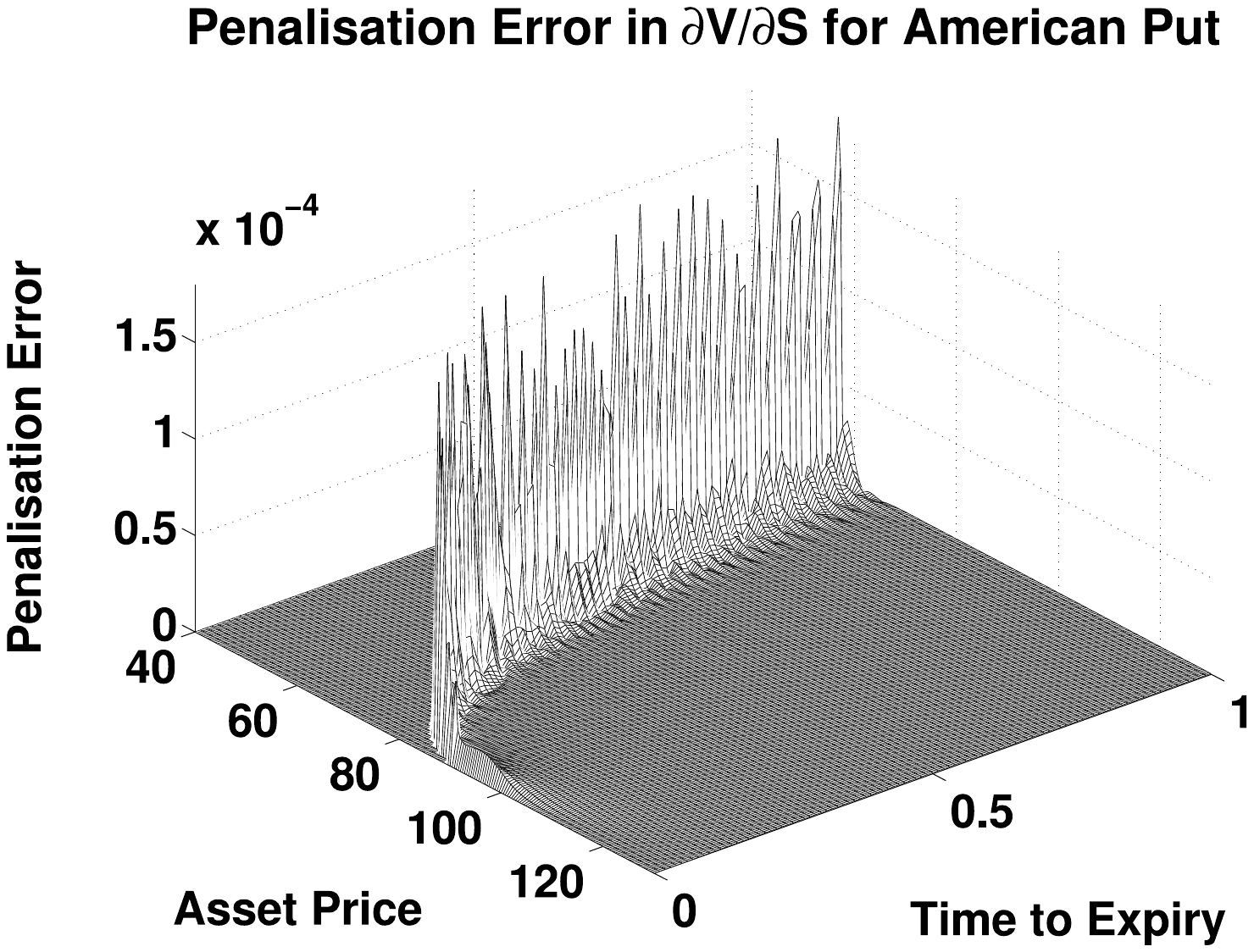}
\caption{Local structure of the penalisation error for the put delta with parameters $\sigma=0.4$, $r=0.05$, and $\epsilon^{-1}=3\cdot 10^4$. The error is largest in  a small layer around the exercise boundary.}
\label{fig:surfacederput}
\end{figure}

Figure \ref{fig:surfacederput} is the corresponding picture for the penalisation error in the first $S$-derivative. The error appears localised in a very narrow region around the exercise boundary.
The jagged shape of the surface results again from an interplay of the penalisation error and discretisation. For the chosen time step and mesh size, 
the width of the region of large error is small compared with the mesh size, and, from one time step to the next, has a different location relative to its nearest grid points. For those time steps, where the location of the maximum is close to a mesh point, the plotted spike is large, whereas if the maximum lies between mesh points, it is small.

\bigskip

\begin{figure}[t]
\centering
\includegraphics[scale=.4]{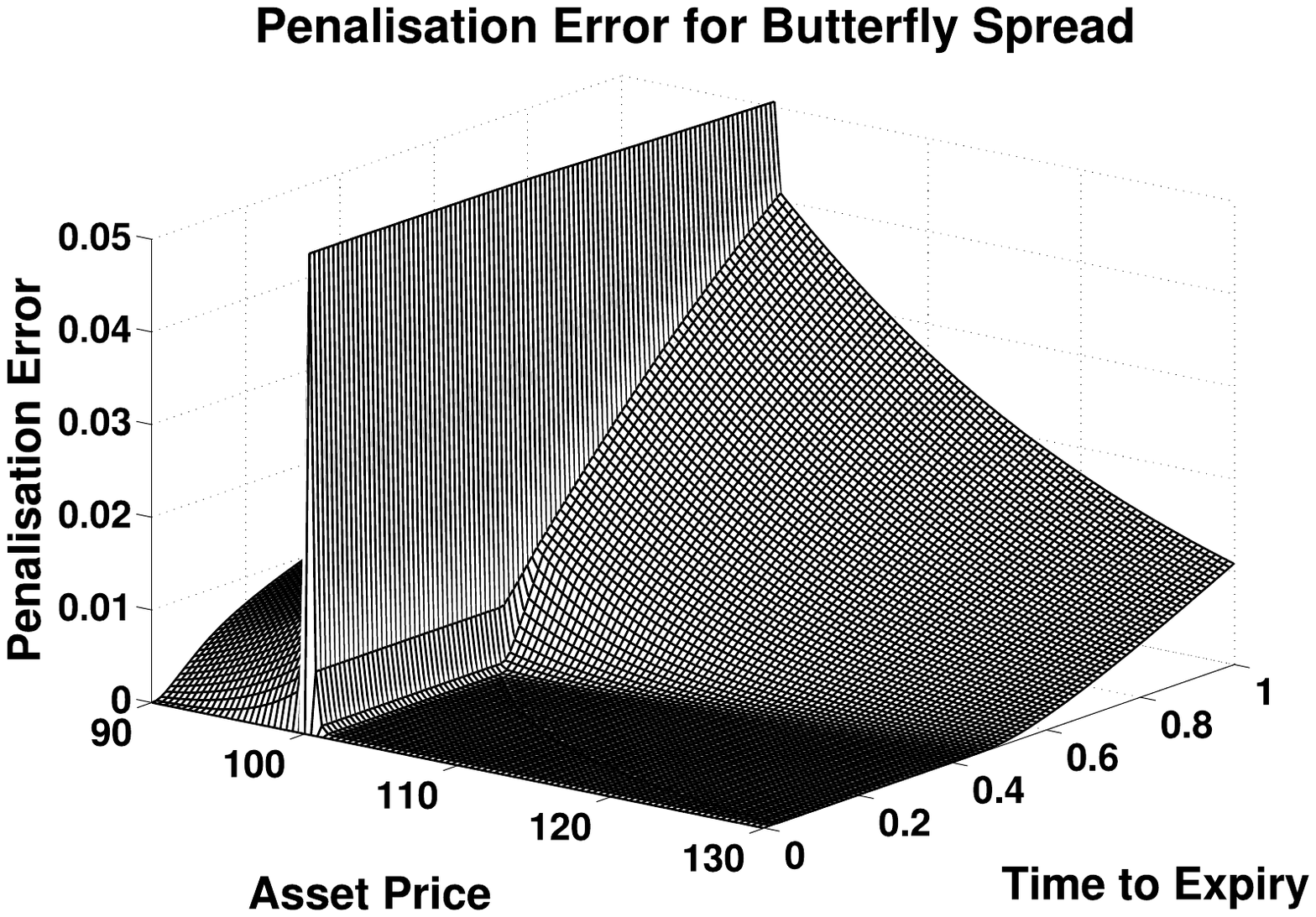}
\caption{Local structure of the penalisation error for the butterfly spread with parameters $\sigma=0.4$, $r=0.05$, and $\epsilon^{-1}=3\cdot 10^4$. The error is largest in a narrow region around the kink of the payoff. It is negligible on the put-like side
up to the point where it is optimal not to exercise the option.}
\label{fig:surfacebfly}
\end{figure}

For the butterfly, as seen from Figure \ref{fig:surfacebfly}, there is an asymmetry in the penalisation error between the call-like side, where the error grows more steeply in time-to-expiry, and the put-like side, where the error is flat up to the point in backward-time where the exercise boundary hits the top of the payoff, and from then on increases for larger time-to-maturity. The error is largest at the strike, constant in time, and decays rapidly on either side.

\bigskip

\begin{figure}[t]
\centering
\includegraphics[scale=.4]{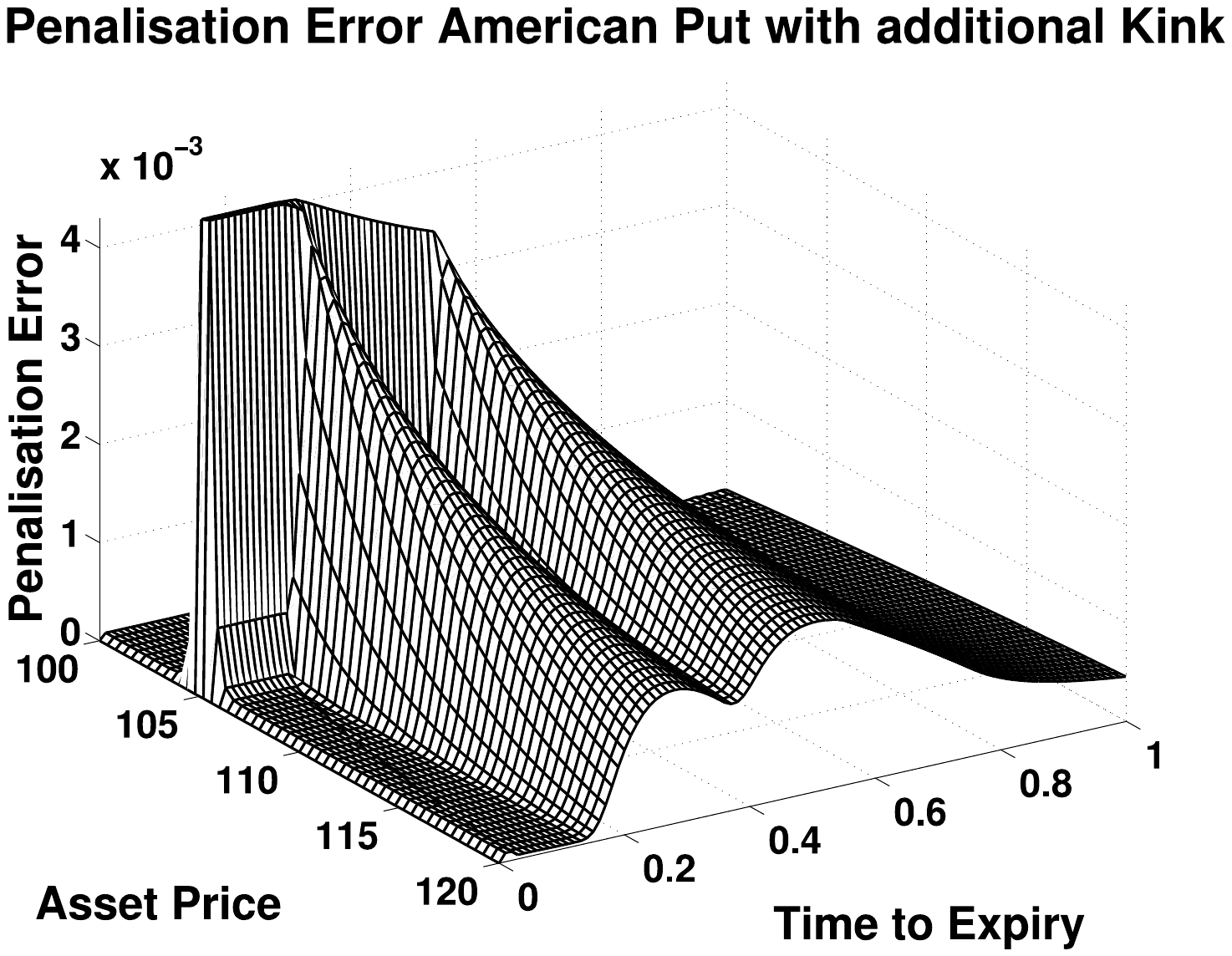}
\caption{Local structure of the penalisation error for the modified put with parameters $\sigma=0.4$, $r=0.05$, and $\epsilon^{-1}=3\cdot 10^4$. It shows a combination of features of the put and butterfly payoff, and a decay in time-to-expiry resulting from the waiting time phenomenon and subsequent
smoothness.}
\label{fig:surfacefunnyput}
\end{figure}

The penalisation error for the modified put is  shown in Figure \ref{fig:surfacefunnyput}. 

\bigskip

Table \ref{Tab:PenErrRates} shows estimated convergence orders of $\Vpen$ to $V$ as $\epsilon\to 0$ for the three payoffs.
We measure spatial errors pointwise in the maximum norm, and similarly for the derivative, which is approximated from the numerical solution by finite differences.
From a sequence of these errors  for small $\epsilon$, we estimate the convergence rate by regression.
Throughout, we use very fine time and space grids to make discretisation errors negligible.

\bigskip

For the standard put, we find results
for the spatial errors in $\Vpen$ and $\partial \Vpen/\partial S$ which appear consistent with
$O(\epsilon)$ and $O(\epsilon^{1/2})$, respectively.
For the butterfly spread, we find $O(\epsilon^{1/2})$ for the spatial error in $\Vpen$\,, but the observed convergence in $\partial \Vpen/\partial S$ is very slow.
Looking at the solutions (cf.\,Figures\,\ref{fig:Put},\,\ref{fig:Butterfly}), one
readily suspects the concave kink of the butterfly spread, which is prevalent in the solution
at all times, to be the reason for the slower convergence. This observation is further supported by
the fact that, for the modified put, we find convergence rates comparable to the rates
of the butterfly spread near expiry, but, further away from expiry, where the concave
kink has been smoothed out (cf.\,Figure\,\ref{fig:FunnyPut}), convergence
improves to roughly $O(\epsilon^{1/2})$ for the spatial errors in $\Vpen$ and $\partial \Vpen/\partial S$.\bigskip

\begin{table}[t]
\begin{tabular}{|c||c|c||c|c||c|c|c|}
\multicolumn{1}{c}{Penalty Approximation} &\multicolumn{2}{c}{Put} & \multicolumn{2}{c}{Butterfly Spread} & \multicolumn{3}{c}{Modified Put}\\
\hline
Time to Expiry & 0.4 & 0.9 & 0.4 & 0.9 & 0.07 & 0.4 & 0.9\\
\hline\hline
Order in $|V-\Vpen|$ & 1.00 & 1.00 & 0.50 & 0.50 & 0.51 & 0.51 & 0.53\\
Order in $|\partial V/\partial S - \partial \Vpen/\partial S|$ & 0.55 & 0.57 & 0.07 & 0.08 & 0.07 & 0.06 & 0.61\\
\hline
\end{tabular}
\caption{
At different points $t$ in time, we measure the convergence rates
in $|V(\cdot,t)-\Vpen(\cdot,t)|$ and $|\partial V/\partial S(\cdot,t) - \partial \Vpen/\partial S(\cdot,t)|$, where $V$ denotes the true solution, as $\epsilon\to 0$. For the put and the butterfly spread, the rates are the same at all times. For the modified put, the convergence rate in $\partial \Vpen/\partial S$ improves hugely when expiry is far into the future
(at which point the free boundary has `overcome' the concave kink).}
\label{Tab:PenErrRates}
\end{table}

In the next section, we will use matched asymptotic expansions for small $\epsilon$ to explain this behaviour and to derive the leading order corrections to the penalty solution.

\section{Approximation by Matched Asymptotic Expansions}
\label{sec:asymp}
\newcommand{\Ge}{\delta}

In this section, we describe the structure of the solutions to the three
canonical problems introduced above, within the framework of perturbation
(asymptotic) analysis and matched asymptotic expansions. 

\bigskip

We exploit the
fact that the penalty parameter $\epsilon$ is `small' and analyse the
problem for any specific value of $\epsilon$ by considering the limit as
$\epsilon \to 0$. The basic idea (see, for example,~\cite{Hinch, Howison05}) 
is to decompose the solution domain into a number of
overlapping regions, whose sizes are related to $\epsilon$, and to
formulate a simplified problem in each of these regions in which some
terms in the equations can be seen \emph{a priori} to be small. The
solutions to the individual problems, which typically contain unknown
functions, are joined together by `matching' in the overlap regions, by
use of Van Dyke's matching principle (see~\cite{Hinch}). This procedure
typically provides the information that is needed to determine any unknown
functions fully. Although the procedure is purely formal, it is confirmed
by the numerical results and, indeed, provides an illuminating
interpretation of the role of error estimates in problems of this kind.

\bigskip

While $\epsilon$ is small, it is a dimensional quantity (with units
of time) and in order to compare different combinations
of the parameters $\sigma$ and $\epsilon$ in a consistent way, we
introduce the small \emph{dimensionless} parameter
\[
\delta =\sigma\epsilon^\frac12
\]
and the limit we consider is $\delta \to 0$.\footnote{
We could have used $r$ or $T$ instead to scale $\Ge$, with the same eventual answer.
The choice we have made makes the intermediate calculations simpler. We also keep writing $\Vpen$ to keep the notation simple.}
We will see later that $\Ge$ is the characteristic width of the `inner' region
between the exercise and hold regions of the option.

\subsection{Asymptotics for Put}
\label{subsec:putasymp}

We first consider an American put option on an asset in the Black-Scholes model with no dividends.
Initial (i.e., near expiry) 
transients are ignored: this means both the penalty term
transient and the American-put transient.
\bigskip

There are three regions. The key region is the `inner' region,
located  around the free  boundary
$S=S^*(t)$ of the American put; it is characterised by an inner variable $x$
defined by
\begin{equation}
\label{VVV}
S =S^*(t)(1+\delta x).
\end{equation}
The choice of scaling in~(\ref{VVV}) is motivated, as in classical
boundary-layer analysis, by the need to retain a
balance between the leading-order terms, including the highest-order derivative,
in the penalty equation
\beq
\mathcal{L}_\mathrm{BS} \Vpen = 
-\frac{1}{\epsilon}\max(K-S-\Vpen,0) =
-\frac{\Gs^2}{\Ge^2}\max(K-S-\Vpen,0),
\label{penaltyeqn}
\eeq
with all remaining terms remaining
smaller as $\delta\to 0$; see (\ref{innerpenalty}) and
(\ref{eqn:matching}). The other two regions are for values of $S$ below and above
the inner region, and are referred to as the outer `exercise' and outer
`hold' regions respectively. 
\bigskip

\begin{figure}
\centering
\includegraphics{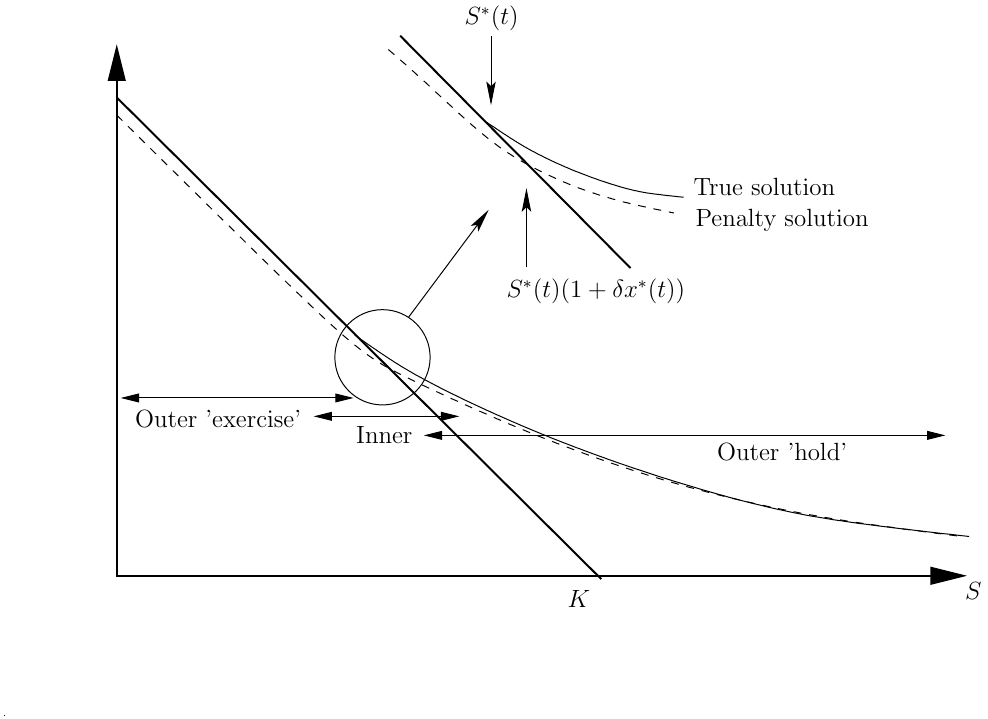}
\caption{Schematic of a put option solution with its
three region structure (outer `hold' and `exercise' regions and an inner region),
and a blow-up of the inner region.
The true solution is the solid curve, the penalty solution is dashed.}
  \label{fig:putscheme}
\end{figure}

The set-up is summarised in Figure~\ref{fig:putscheme},
and a summary of the expansions we find is given at the end of this
subsection.


First, write
\begin{equation}
\label{defWpen}
\Vpen(S,t)=K-S+\Wpen(S,t)
\end{equation}
and expand 
\begin{equation}
\label{outexpand}
\Wpen(S,t)\sim W_0(S,t)+\Ge W_1(S,t)+\Ge^2W_2(S,t)+\cdots.
\end{equation}
We expect $W_0(S,t)=P(S,t)-(K-S)$, where $P$ is the true put value, and $W_1(S,t)=0$ (because smooth pasting
always leads to a smaller error than a barrier-type `pinned' condition, cf.\ the asymptotic results on Bermudan options and discrete
barrier options in \cite{howisonsteinberg07,howison07}).

\subsubsection{Outer `hold' region $S>\Sst$}

$W_0$ satisfies $\mathcal{L}_{BS} W_0=-rK$ and all $W_i$ for $i>0$ satisfy the homogeneous Black-Scholes PDE, because the penalty is not active. 
Then Taylor-expanding $W(S,t)$ about $S=\Sst$ and writing the result
in terms of $x$ gives the outer expansion expanded in inner variables
as
\begin{align}
\Wpen(S,t)  =& \;\; W(S^*(t)(1+\Ge x),t) \notag \\
\sim &\;\; W_0^*(t)  \notag \\
& +\Ge\left(x\Sst W_{0S}^*(t) + W_1^*(t)\right)\notag \\
& +\Ge^2 \left(\harf x^2 \Sst^2 W_{0SS}^*(t) +x \Sst W_{1S}^*(t) +
  W_2^*(t)\right)\notag \\
&+\cdots,
\label{outerinner}
\end{align}
as $\Ge\rightarrow 0$, where
\[
W_0^*(t) = W_0(\Sst,t), \qquad W_{0S}^*(t) = \pda{W_0}{S}(\Sst,t)\quad \text{etc.}
\]
are functions of $t$ alone and as yet unknown. Hence, in the absence of spatial boundary conditions, we can do no more in this region for now.

\subsubsection{Outer `exercise' region $S<\Sst$}
\label{subsubsec:outex}

This is the region below the exercise point, in which the penalty term is active. The penalty equation (\ref{penaltyeqn})
becomes
\[
\mathcal{L}_\mathrm{BS} \Wpen = rK +\frac{\Gs^2}{\Ge^2}\Wpen.
\]
Inserting the expansion (\ref{outexpand}) and matching individual powers of $\Ge$ gives
\begin{align*}
W_0&= 0 &\qquad\text{(coefficient of $\Ge^{-2}$)},\\ 
W_1 &= 0 &\qquad\text{(coefficient of $\Ge^{-1}$)},\\ 
r K + \sigma^2 W_2 &= 0 &\qquad\text{(coefficient of $\Ge^{0}$)},\\ 
W_3 &= 0 &\qquad\text{(coefficient of $\Ge^{1}$)},
\end{align*}
so we obtain
\begin{equation}
\label{outersmall}
\Wpen \sim - \Ge^2 rK/\Gs^2 +O(\Ge^4).
\end{equation}
This dictates the scaling of $W$ in the inner region.

\subsubsection{Inner region}

Make the change of variables to $x$ and $t$, and expand
\begin{align}
\Wpen(S,t)&=\wpen(x,t)\\
& \sim w_0(x,t) + \Ge w_1(x,t) + \Ge^2w_2(x,t)+\cdots.
\label{innerexpansion}
\end{align}    
The penalty equation becomes
\begin{align}
\pda{\wpen}{t} -\frac{\dot{S}^*}{\Ge S^*}(1+\Ge x)\pda{\wpen}{x}
+\harf \Gs^2 \frac{(1+\Ge x)^2}{\Ge^2}\pdb{\wpen}{x}
+&r\frac{1+\Ge x}{\Ge}\pda{\wpen}{x}  - r \wpen \notag \\
& = rK + \begin{cases} 0 &\qquad x>x^*,  \\ 
\frac{\Gs^2}{\Ge^2}\wpen &\qquad
  x<x^*,
\end{cases}
\label{innerpenalty}
\end{align}
where $\dot{S}^*=\romd S^*/\romd t$, and 
with $\wpen=0$ and $\prt \wpen/\prt x$ continuous at $x=x^*$ (i.e.,
$x^*$ is the crossing point of the penalty solution).

\subsubsection{Matching} As $x\to -\infty$,  we have
(cf.\ (\ref{outersmall}))
\[
w_0\to 0, \qquad w_1\to 0, \qquad w_2\to -rK/\Gs^2,
\]
and as $x\to +\infty$ we have (compare~\eq{outerinner})
\begin{align*}
w_0(x,t)  &\sim W_0^*(t) + o(1), \\
w_1(x,t) &\sim x\Sst W_{0S}^*(t)+W_1^*(t) +o(1),\\
w_2(x,t)& \sim \harf x^2 \Sst^2 W_{0SS}^*(t)+ x\Sst W_{1S}^*(t)
+W_2^*(t) +o(1).
\end{align*}

The largest terms in~\eq{innerpenalty} are $O(1/\Ge^2)$. When we
substitute the expansion (\ref{innerexpansion}) in and collect terms, we get, at $O(1/\Ge^2)$,
\begin{equation}
\harf \Gs^2\pdb{w_0}{x}
 = \begin{cases} 0 &\qquad x>x^*,  \\ 
\Gs^2w_0 &\qquad
  x<x^*,
\end{cases}
\label{eqn:matching}
\end{equation}
and the only solution that vanishes at $x=-\infty$, has continuous
first derivative at $x=x^*$, and tends to a constant at $x=+\infty$, is
$w_0(x,t)\equiv 0$. This tells us that 
\[
W_0^*(t) =0
\]
as expected. Because $W_0(S,t)$ is the difference between the vanilla
value of the put and the payoff, its $S$-derivative vanishes at
$S=\Sst$ --- this is smooth pasting. (In more detail, because
$W_0(S,t)$ has the right value at $S=\Sst$ and the right payoff,
uniqueness for solutions of the BSPDE in a  parabolic domain tells us
that it \emph{is} the vanilla put value.) Hence, $W_{0S}^*(t)=0$.
Now, at $O(1/\Ge)$ in~\eq{innerpenalty}, we get
\[
\harf \Gs^2\pdb{w_1}{x}
 = \begin{cases} 0 &\qquad x>x^*,  \\ 
\Gs^2w_1 &\qquad
  x<x^*.
\end{cases}
\]
As $W_{0S}^*(t)=0$, $w_1(x,t)$ has no linear term at $x=+\infty$, and so,
by the same argument as above, it vanishes too, confirming that the inner
scaling for $W$ is indeed $O(\Ge^2)$. Hence, $W_1^*(t)=0$, and we can return to
the outer region $S>\Sst$ to show that $W_1(S,t)\equiv 0$ (zero
payoff, zero value on $S=\Sst$).
Now we come to the first non-trivial term. 
At $O(1)$ in~\eq{innerpenalty}, we have 
\[
\harf \Gs^2\pdb{w_2}{x}
 = rK + \begin{cases} 0 &\qquad x>x^*,  \\ 
\Gs^2w_0 &\qquad
  x<x^*.
\end{cases} 
\]
For $x<x^*$, the solution that tends to $-rK/\Gs^2$ at $-\infty$ and vanishes
at $x=x^*$ is 
\beq
\label{rinsey0}
w^-_2(x,t) = rK\left(\rome^{\sqrt{2}(x-\xst)}-1\right)/\Gs^2.
\eeq
For $x>x^*$, the solution that vanishes at $x=x^*$ and whose derivative matches
(\ref{rinsey0}) is
\beq
\label{rinsey1}
w_2^+(x,t)= \frac{rK}{\sigma^2}(x-\xst)^2 + rK\sqrt{2}/\sigma^2 (x-\xst).
\eeq

\bigskip

Now comes the key point. From the matching, we now know that
\[
w_2^+(x,t)\sim \harf x^2 \Sst^2 W_{0SS}^*(t) + W_2^*(t), \qquad x\to\infty.
\]
There is no linear term because $W_1(S,t)=0$. Comparing
with~\eq{rinsey1}, we find that
\begin{align*}
rK/\Gs^2 &= \harf \Sst^2 W_{0SS}^*(t) &\qquad\text{(coefficient of
$x^2$)},\\ 
-2rK\xst/\Gs^2 + rK\sqrt{2}/\Gs^2&= 0 &\qquad\text{(coefficient of
$x$)},\\ 
rK{x^*}(t)^2/\Gs^2 -\sqrt{2}rK\xst/\Gs^2&= W_{2}^*(t) &\qquad\text{(constant
  coefficient)}.
\end{align*}
The first of these confirms the boundary Gamma of the vanilla put.  
The second gives
\[
\xst= 1/\sqrt{2}.
\]
The third gives
\[
W_2^*(t)=-\harf rK/\Gs^2.
\]

In original variables, the crossing point is at
\begin{align*}
S^{\epsilon}(t) &= \Sst(1+\Ge \xst + \ldots) \\
& = \Sst\left(1 + \sqrt{\epsilon} \Gs /\sqrt{2} + \ldots \right)
\end{align*}
as $\Ge^2 = \Gs^2 \epsilon$, and the boundary value of the correction is, at leading order,
\[
\Ge^2W_{2}^*(t) = -\harf rK \epsilon.
\]

\subsubsection{Summary of results and numerical verification}

In summary, the penalisation error for the exercise boundary is
\begin{eqnarray}
\label{asympfb}
\Sst-S^\epsilon(t) = {\small \frac{1}{\sqrt{2}}} \sigma \epsilon^{1/2} +
o(\epsilon^{1/2}),
\end{eqnarray}
and for the penalised value function $\Vpen$, compared to the true solution $P$
for the put,
\begin{eqnarray}
\label{asympsumm}
&& \\ \nonumber
(P-\Vpen)(S,t)
=
\epsilon \, r K \;
\left\{
\begin{array}{rl}
 1
& \quad S<\Sst (1-O(\sigma \epsilon^{1/2})) \\
W^-(S,t)  &
\quad \Sst (1-O(\sigma \epsilon^{1/2})) < S < \Sst \\
W^-(S,t) + \frac{W_0(S,t)}{\epsilon r K} &
\quad \Sst < S < S^\epsilon(t) \\
W^+(S,t) + \frac{W_0(S,t)}{\epsilon r K} & 
\quad {S^\epsilon(t) < S < S^\epsilon (1+ O(\sigma \epsilon^{1/2}))} \\
\harf D(S,t) 
& \quad S>\Sst (1+O(\sigma \epsilon^{1/2}))
\end{array}
\right\}
+  o(\epsilon),
\end{eqnarray}
where
\begin{eqnarray*}
W^-(S,t) &=& 1-{\rome}^{\sqrt{2} (S-S^\epsilon(t))/(\sigma \epsilon^{1/2})}, \\
W^+(S,t) &=& \harf - \frac{(S-\Sst)^2}{\sigma^2 \epsilon},
\end{eqnarray*}
and $W_0$ from earlier. Note that for the relevant range $0<S-\Sst=O(\sigma\epsilon^{1/2})$,
\begin{eqnarray*}
W_0(S,t) &=& P(S,t)-(K-S) = \harf (S-\Sst)^2  P_{SS}(\Sst,t) + o(\sigma^2\epsilon) =  O(\sigma^2 \epsilon)
\end{eqnarray*}
due to smooth pasting, and $W^-$ and $W^+$ are $O(1)$ in their relevant ranges.
Finally, $D$ is defined in the hold region, with $0\le D\le 1$.
More precisely, the function $D$ satisfies the Black-Scholes PDE as the penalty term is not active.
It is 1 at the exercise boundary, $D(\Sst,t) =1$, and 0 at maturity, $D(S,T)=0$. Hence, it is interpretable as the value of an option with zero payoff at maturity which pays a fixed amount of 1 when (if) the stock crosses $\Sst$ from above.
The first-order correction to the value is independent of $\Gs$ to leading order.
\bigskip 

Interestingly, the continuity
correction for a Bermudan option (i.e., the difference to the American option)
found in \cite{howison07} has the same boundary value if we set $\delta^2 = \sigma^2 \Delta T /2$, where $\Delta T$ is the interval between exercise dates.
\bigskip

Table \ref{Tab:PenErrComp} compares the corrections based on 
the leading terms in (\ref{asympfb}) and (\ref{asympsumm})
against the numerically computed penalisation error and finds excellent agreement.
\begin{table}[t]
\begin{tabular}{|c||c|c|c|c|}
\multicolumn{1}{c}{Penalisation Error} &\multicolumn{1}{c}{Computed} & \multicolumn{1}{c}{Predicted} & \multicolumn{1}{c}{Relative Difference} \\
\hline
Value in Exercise Region & 4.9975e-04 & 5.0000e-04 & -5.0075e-04 \\
Value in Hold Region & 2.5070e-04 & 2.5000e-04 & 0.0028 \\
Exercise Boundary & 0.0174 & 0.0165 & 0.0516 \\
\hline
\end{tabular}
\caption{For the American put, the maximum penalisation error in exercise and hold regions separately, and the error of the exercise boundary, for
$\epsilon^{-1} = 100$, $\sigma=0.4$, $r=0.05$, $K=100$, $T=1$, hence $\delta= 0.04$. 
The numerical result is compared with
the first order correction from the asymptotic analysis, as summarised in (\ref{asympsumm}) and (\ref{asympfb}).}
\label{Tab:PenErrComp}
\end{table}

\subsection{Butterfly Spread}
\label{subsec:butterfly}


Still in the Black-Scholes framework without dividends, consider now a butterfly spread with payoff
\[
\begin{cases} 
\max(V_0 +\alpha_1(S-K),0) \quad & S<K,\\
\max(V_0 + \alpha_2(K-S),0) \quad & S> K,
\end{cases} 
\]
where $V_0$, $\alpha_1$, $\alpha_2$ are all positive.
Denote again the penalty value by $\Vpen(S,t)$, the true value by $B(S,t)$.
Consider the situation where $B(S,t)$ has a free boundary on the put-like
bit of the payoff ($S>K$) but on the call-like bit, the  value of
$B(K,t)$ is anchored to $V_0$. This certainly happens for short times
before expiry (as the Black-Scholes operator on the payoff is positive).
See also Figure \ref{fig:Butterfly}.
The former put-like bit is analysed as before so we focus on the region around the convex kink at $S=K$.

\begin{figure}
\centering
\includegraphics{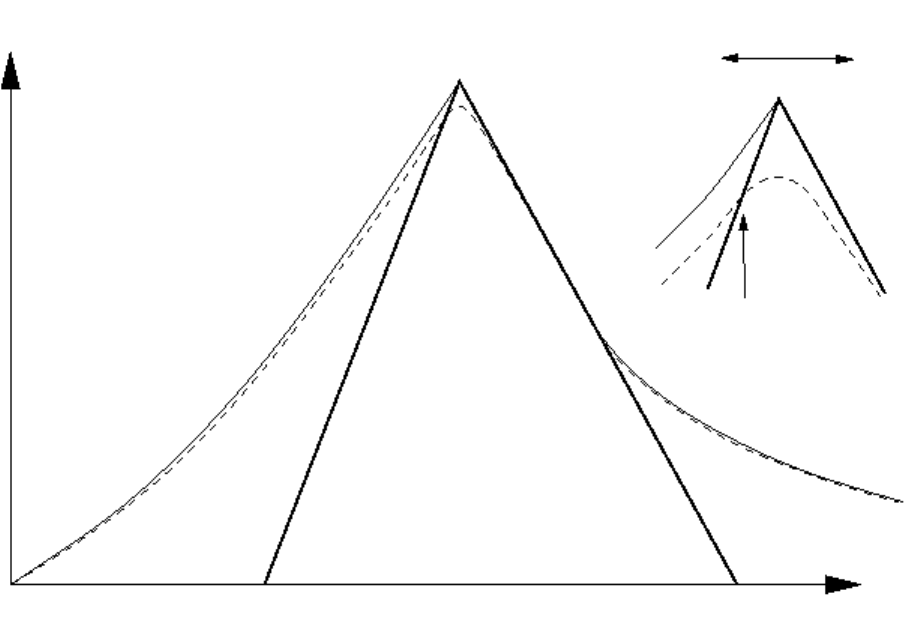}
\caption{Schematic of a butterfly option solution and blow-up of the inner
  region near the peak. The true solution is the solid curve, the penalty solution is dashed.}
\label{fig:blfyscheme}
\end{figure}

\subsubsection{Outer `hold' region $S<K$}
The inner variable is now $S=K(1+\Ge x)$,
the outer solution for $S<K$ is of the form
\[
\Vpen(S,t) = B(S,t)+\Ge V_1(S,t)+\cdots
\]
and its inner expansion near $S=K$ is
\[
V_0 +\Ge(xB_S^*(t)+V_1^*(t))+\cdots,
\]
where $B_S^*(t) = \lim_{S\uparrow K} \prt B\!\!\;/\!\!\;\prt S\,(S,t)$ and $V_1^*(t)=V_1(K,t)$. 

\subsubsection{Outer `exercise' region $S>K$}

This is identical to the case of the put in Section \ref{subsubsec:outex}.

\subsubsection{Inner region}
The payoff in inner variables is
\[
\begin{cases} V_0 +\Ge \alpha_1 Kx \quad & x<0,\\
              V_0 - \Ge \alpha_2 Kx \quad & x> 0.
\end{cases} 
\]
This suggests that the inner solution is of size $O(\Ge)$, not
$O(\Ge^2)$, and this is consistent with the left-hand 
outer solution meeting the
payoff at an angle (not smooth pasting).
Write the inner expansion in the form
\[
\Vpen(S,t)\sim V_0 +\Ge v_1(x,t)+\cdots.
\]
Also let $x^*$ (which is negative) be the point at which the penalty
solution crosses the payoff. The leading order inner equation is
\[
\harf\Gs^2\pdb{v_1}{x} = \Gs^2  
\begin{cases} v_1 - \alpha_1 Kx \quad & x<0,\\
              v_1 +  \alpha_2 Kx \quad & x> 0.
\end{cases} 
\]

\subsubsection{Matching}
The solution is $C^1$ at both $x=0$ and $x=x^*$, where $C^1$ is the space of continuously differentiable functions. As $x\to\infty$,
$v_1\sim -\Ga_2 Kx$ because the solution is accurate to $O(\Ge^2)$ in the
`exercise'
region to the right of $S=K$. So,
\[
v_1=
\begin{cases}   \alpha_1 Kx +
  a(t)K\cosh(x\sqrt{2})+b(t)K\sinh(x\sqrt{2})
\quad & x<0,\\
               - \alpha_2 Kx + a(t)K\rome^{-x\sqrt{2}} \quad & x> 0,
\end{cases} 
\]
for some $a(t)$, $b(t)$. This is continuous at $x=0$, and  
continuity of $\prt v_1/\prt x$ at $x=0$ gives
\beq
\label{rinsey11}
\Ga_1 +b(t)\sqrt{2}=-\Ga_2 -a(t)\sqrt{2}.
\eeq
Now, at $x=x^*$, $v_1$ meets the payoff and joins onto the outer solution: 
\[
v_1(\xst,t)= \Ga_1 K \xst, \qquad \pda{v_1}{x}= B_{S}^*(t),
\]
from which
\begin{gather}
\Ga_1 \xst + a(t) \cosh(\xst\sqrt{2}) +b(t)\sinh(\xst\sqrt{2}) 
= \Ga_1\xst,\label{home1}\\
\Ga_1+\sqrt{2}\left(a(t)\sinh(\xst\sqrt{2}) +
    b(t)\cosh(\xst\sqrt{2}\right)= B_\Sst. \label{home2}
\end{gather}  
This, with~\eq{rinsey11}, is three equations for $a(t)$, $b(t)$ and $\xst$. 

From these, we readily find that
\[
\xst=-\frac{1}{\sqrt{2}}\log\frac{\Ga_1+\Ga_2}{\Ga_1-B_\Sst}.
\]
This is clearly negative since we have $0<B_S^*<\Ga_1$. Also, it
tends to $-\infty$ as $B_S^*\to\Ga_1$ from below -- that is, as (if)
the free boundary moves away from $S=K$.

\subsubsection{Summary of results}

In the scenario where there is a non-trivial exercise boundary $\Sst>K$,
and the solution is pinned to the payoff at $S=K$,
we have computed the crossing point of the penalty solution just left of the 
strike.
From this we derive the correction term $V_1(K,t)=\Ga_1K\xst$,
although not explicitly because of the complicated time dependence
of $\xst$ via $B_\Sst$.


\subsection{American Put under Jump-Diffusion}


We now extend the analysis under Black--Scholes 
to include jumps of relative size $J$, at the jump time of a compound Poisson process with rate $\lambda$.
We only do this for the standard put payoff to illustrate the extensions over Black-Scholes. The results are qualitatively very similar to the Black-Scholes case and the analysis suggest this will also be the case for other payoffs.
In addition, we also account for continuously paid proportional dividends of rate $q$, where we assume $q\le r$. (The solution for $q > r$ is qualitatively different.)\bigskip

The penalised equation is similar to (\ref{penaltyeqn}), specifically, for the put under a jump model
\beq
\mathcal{L}_\mathrm{BSJ}\Vpen = -\frac{1}{\epsilon}\max(K-S-\Vpen,0) = -\frac{\Gs^2}{\Ge^2}\max(K-S-\Vpen,0),
\label{penaltyeqnJ}
\eeq
 where $\mathcal{L}_\mathrm{BSJ}$ is defined in (\ref{BSJOP}).
Smooth pasting still holds for the vanilla American put value $P(S,t)$, and applying
these conditions just to the right of the exercise boundary now gives the boundary Gamma as
\beq
\label{FB1J}
\left.\pdb{P}{S}\right|_{S\downarrow\Sst}\!\! = \frac{2}{\Gs^2 \Sst^2} \left(r K - (q+\omega\lambda) \Sst - \lambda \EE[P(J \Sst,t)-P(\Sst,t)]\right)
\equiv \frac{2}{{\Sst}^2} \constgam,
\eeq
which we use to define the function $\constgam$ for future reference.
Note that $\constgam$ depends on $P(S,t)$ for all $S>0$ via the term $\EE[P(J \Sst,t)-P(\Sst,t)]$. 
\bigskip

There are three regions as in Section \ref{subsec:putasymp}, see particularly Figure \ref{fig:putscheme}.

\subsubsection{Outer `hold' region $S>\Sst$}

It will be useful to introduce again $\Wpen(S,t)$ as in (\ref{defWpen}).
As $S \rome^{-q(T-t)}$ and $K \rome^{-r(T-t)}$ satisfy $\mathcal{L}_{BSJ} V=0$ individually
(in fact, their difference is the value of a forward), one gets
\[
\mathcal{L}_\mathrm{BSJ} \Vpen(S,t) =
\mathcal{L}_\mathrm{BSJ}  \Wpen(S,t) - 
\mathcal{L}_\mathrm{BSJ} (S-K) = 
\mathcal{L}_\mathrm{BSJ}  \Wpen(S,t) - qS + rK.
\]

\subsubsection{Outer `exercise' region $S<\Sst$}

With the above substitution, as $W(S,t)\le 0$ in this region by assumption,
\[
\mathcal{L}_\mathrm{BSJ} \Wpen(S,t) = rK - qS + \frac{\sigma^2}{\Ge^2} \Wpen(S,t).
\]
Inserting the expansion for $\Wpen(S,t)$, $W_0(S,t)$ and $W_1(S,t)$ vanish in the outer
region $S<\Sst$ as before; however, when determining $W_2(S,t)$ in the
`exercise' region, we have to account for jumps into the other regions, particularly into the outer `hold' region $S>\Sst$, where $\Wpen(S,t)$ will not be small. Thus, at $O(1)$,
\[
\lambda \EE[W_0(JS,t)-W_0(S,t)] = rK-qS + \sigma^2 W_2(S,t),
\]
and, solving for $W_2(S,t)$,
\begin{eqnarray}
\nonumber
W_2(S,t) &=& - \left( (rK-qS) - \lambda \EE[P(SJ,t) - (K-JS)] \right)/\sigma^2 \\
\nonumber
&=& - \left( (rK-qS) - \lambda \EE[P(SJ,t) - P(S,t) + (J-1) S] \right)/\sigma^2.
\label{defW2}
\end{eqnarray}

\subsubsection{Inner region}

Similar to (\ref{innerpenalty}), we now have
\begin{multline}
\pda{\wpen}{t} -\frac{\dot{S}^*}{\Ge S^*}(1+\Ge x)\pda{\wpen}{x}
+\harf \Gs^2 \frac{(1+\Ge x)^2}{\Ge^2}\pdb{\wpen}{x}
+r\frac{1+\Ge x}{\Ge}\pda{\wpen}{x}  - r \wpen\\
+ \; \EE[\Wpen(J S^* (1+\Ge x),t) - \Wpen(S^* (1+\Ge x),t)]\\
= rK - q S^*
(1+\Ge x) + \begin{cases} 0 & x>x^*,\\ 
\frac{\Gs^2}{\Ge^2}\wpen &
  x<x^*.
\end{cases}
\label{innerpenaltyJ}
\end{multline}

The non-local term is written in terms of the outer solutions because it acts on the scale of the outer variables.
Writing the expectation term as integral and expanding in $\Ge$ gives, at leading order,
\[
\Ge S^* x  \int_0^\infty  J \, \Wpen_S(J S^*,t) \, g(J) \, \text{d}J.
\]
The simple expansion has a natural interpretation: given a jump size, all jumps starting from the inner region and ending in
the outer region end up close to each other. If we were going to a higher
order of accuracy (which we are not), we would have to treat the small
jumps -- those which both start and end in the inner region -- separately.
So the integral for the expectation would have its range split into inner
and outer parts, and so on.\bigskip

Comparing terms $O(1/\Ge^2)$ and $O(1/\Ge)$ gives again that $w_0(x,t)$ and $w_1(x,t)$ vanish, and now, at $O(1)$,
\[
\harf \Gs^2\pdb{w_2}{x}
 = \Gs^2 \constgam + \begin{cases} 0 &\qquad x>x^*,  \\ 
\Gs^2w_2 &\qquad
  x<x^*.
\end{cases} 
\]

\subsubsection{Matching}

First, we match the inner solution with the outer solution in the exercise region.
The matching of $w_2^-$ for $x\rightarrow -\infty$ is now to a non-constant value,
but it is clear that $W_2(S,t)$ from (\ref{defW2}) approaches $\constgam$ for $S\rightarrow S^*$ in the
`outer' variables, and matching in an overlap region demands that, as $x\rightarrow -\infty$,
$w_2(x,t) \rightarrow -\constgam$.
Then calculations identical to before give,
for $x<x^*$,
\[
w^-_2(x,t) = \constgam \left(\rome^{\sqrt{2}(x-\xst)}-1\right), 
\] 
and, for $x>x^*$,
\beq
w_2^+(x,t)= \constgam \left((x-\xst)^2 + \sqrt{2}(x-\xst)\right).
\eeq

Matching with the outer region $S>S^*$ as before gives
\begin{eqnarray*}
\constgam &=& \frac{1}{2} \Sst^2 W^*_{0SS}(t), \\
\constgam (-2 \xst+\sqrt{2}) &=& 0, \\
\constgam (\xst^2 - \sqrt{2} \xst) &=& W_2^*(t).
\end{eqnarray*}

The first equation recovers the jump diffusion gamma from earlier.
Interestingly, the relative position $x^*$ of the penalty crossing point in relation to the exercise boundary, which is given by the second equation, is unaffected by the jumps.
The last equation, upon inserting $x^*$, shows again that the penalisation error at the free boundary is half the
value one would get by extrapolation from the outer exercise region.

\bigskip

The penalisation error of the exercise boundary is the same in the presence of jumps as in the Black-Scholes model.
For the value function, the first-order correction to the value is again independent of $\Gs$.

\subsection{Discussion of Results}

We now return to discuss the results summarised earlier in
Table \ref{Tab:PenErrRatesSumm} in the light of the findings of this section.

\bigskip

The lack of uniform convergence
of the penalty butterfly Delta, denoted by the `$\star$', results
from the jump of the exact Delta at the strike, which cannot
be matched simultaneously on both sides by the continuous penalty Delta.
However, the asymptotic analysis also reveals that the error in the Delta is
$O(\epsilon^{1/2})$ except in a region which is of width $O(\epsilon^{1/2})$.

\bigskip

The rates in the $H^1$ norm can be explained by the asymptotic analysis as follows:
for the put (as example of a convex payoff), we have an error in the derivative
of $O(\sqrteps)$ in the inner region of width $O(\sqrteps)$, resulting in an $L_2$ error of the derivative of
$$\sqrt{O\big((\sqrteps)^2 \sqrteps\big)} = O(\epsilon^{3/4}).$$ The error in the derivative in the outer region is integrable and $O(\epsilon)$ 
(we can just differentiate the outer expansion) and therefore negligible.
The contribution of the 
zero order term in the $H^1$ error is also of order 1.
A similar argument explains the order $1/4$ for the butterfly.

\section{General Upper and Lower Value Bounds}
\label{sec:comparison}

In the previous section, we computed the penalisation error to leading order in the penalty parameter, and noted a distinct difference in the error
for the put, which $O(\epsilon)$, and a butterfly payoff, which is $O(\epsilon^{1/2})$.
We now show that a distinction into categories of piecewise smooth payoffs with convex and non-convex kinks allows us to derive general upper and lower bounds on the value function.
Under the location of a `convex kink' of a continuous, piecewise smooth function $\Psi$ we understand a point $\bar{S}$ where
$\Psi'(\bar{S}-) \equiv \lim_{S\uparrow \bar{S}} \Psi'(S) < \lim_{S\downarrow \bar{S}} \Psi'(S) \equiv\Psi'(\bar{S}+)$, and similarly for concave kinks.
We work under jump-diffusion models.

\subsection{A Maximum Principle Argument}

Considering the penalised equation
\beq
\label{penjump}
-\mathcal{L}_\mathrm{BSJ}\Vpen = \frac{1}{\epsilon}\max(\Psi-\Vpen,0),
\eeq
it is automatically true that $-\mathcal{L}_\mathrm{BSJ} \Vpen \ge 0$, and if (where) $\Vpen>\Psi$, then $\mathcal{L}_\mathrm{BSJ} \Vpen = 0$, 
such that a complementarity condition is satisfied
and $\min(-\mathcal{L}_\mathrm{BSJ} \Vpen,\Vpen-\Psi)\le 0$. Hence, $\Vpen$ only fails to be a solution to
\[
\min(-\mathcal{L}_\mathrm{BSJ} \Vpen,\Vpen-\Psi) = 0
\]
where $\Vpen\ge \Psi$ is violated.
\bigskip

We begin with an elementary analysis of this latter inequality constraint.
Consider $\Wpen=\Vpen-\Psi$, such that the biggest violation of $\Vpen\ge \Psi$
is given at a global negative minimum of $\Wpen$ (if one is attained).
Note that, for $t<T$, the solution $\Vpen$ to (\ref{penjump}) is twice continuously differentiable everywhere in $S$, by standard regularity arguments.
We first consider points at which $\Psi$ is also smooth, i.e., excluding kinks.
Then, at any such negative minimum $S$ of $\Wpen$, by inspection of the individual terms,
\begin{multline}
\mathcal{L}_\mathrm{BSJ} \Wpen
= \pda{\Wpen}{t} + \harf\Gs^2S^2\pdb{\Wpen}{S}
+(r-q-\omega \lambda) S\pda{\Wpen}{S}-r\Wpen\\
+ \lambda\EE[\Wpen(JS,t)-\Wpen(S,t)] > 0.
\end{multline}
From
\[
- \frac{1}{\epsilon} (\Psi-\Vpen) = \mathcal{L}_\mathrm{BSJ} \Vpen = 
\mathcal{L}_\mathrm{BSJ} \Wpen + \mathcal{L}_\mathrm{BSJ} \Psi
\]
it follows that
\begin{equation}
\label{maxbnd}
\Vpen > \Psi + \epsilon \, \mathcal{L}_\mathrm{BSJ} \Psi.
\end{equation}
For piecewise linear payoffs $\Psi$, it is straighforward to show that, again excluding kinks,
$\mathcal{L}_\mathrm{BSJ} \Psi$ is bounded from below, uniformly for all $S$.
Also, $\Wpen=\Vpen-\Psi$ does not have any negative minima at convex kinks of $\Psi$, i.e.,
points with $\Psi'(S-)<\Psi'(S+)$.\bigskip

Summarising, the biggest violation of the inequality $\Vpen\ge \Psi$ is either
bounded by the maximum of $\epsilon \mathcal{L}_\mathrm{BSJ} \Psi$ taken over the smooth intervals of $\Psi$, 
or attained at concave kinks
or at one of the boundaries $S=0$ or $S\rightarrow\infty$.
We will come back to this observation later to obtain easily computable bounds on the solution.

\subsection{Constructing Bounds on the Value Function}

To treat the solution uniformly in the hold and exercise regions, inclusive of kinks,
it is convenient to work in the framework of \emph{viscosity solutions}.
We use the equation
\begin{equation}
\label{lcplog}
\min(-\mathcal{L} u, u-\psi) = 0
\end{equation}
written in $\log$ coordinates on $\mathbb{R}$, with $\mathcal{L}$ as in (\ref{jumplogoperator}), $\psi(x)=\Psi(S)=\Psi(S_0 \exp(x))$, 
and its penalised version
\begin{equation}
\label{lcplogpen}
-\mathcal{L} \upen = \frac{1}{\epsilon} \max(\psi-\upen,0).
\end{equation}
This avoids technicalities of boundary conditions
and hence discontinuous viscosity solutions, and we can use
the definition from \cite{Pham_OptStop_JumpDiff}, which we tailor slightly to our setting for convenience:
\begin{definition}[Viscosity Solution]
$u\in C([0,T]\times\mathbb{R})$ is a \emph{viscosity supersolution (subsolution)} of (\ref{lcplog}), if
\[
\min(-(\mathcal{L} \phi)(x,t), \phi(x,t)-\psi(x)) \ge 0 \quad (\le 0)
\]
whenever $\phi \in C^2([0,T]\times\mathbb{R})\cap C_2([0,T]\times\mathbb{R})$ and $u-\phi$ has a global minimum
(maximum) at $(x,t) \in [0,T)\times \mathbb{R}$ with $v(x,t)=\phi(x,t)$.
$u$ is a \emph{viscosity solution} iff it is a super- and subsolution.
\end{definition}
Here, $C^2$ is the space of twice continuously differentiable functions, and $C_2$ the space of continuous functions with at most quadratic growth at $\infty$.
This includes the put and butterfly payoffs, but \emph{not} the call payoff in $\log$-coordinates.
It is clear, though, that the results can be extended (e.g., by a coordinate transformation identical to the logarithm for small values, the identity for large values, and a smoothly increasing transition in between).
We further assume that the density $\nu$ has bounded third moments.

\bigskip
Pham \cite{Pham_OptStop_JumpDiff} shows that under these conditions (\ref{lcp}) satisfies a comparison principle.
\begin{theorem}[Theorem 4.1 in \cite{Pham_OptStop_JumpDiff}]
\label{theo:phamcomp}
If $u$ and $v$ are uniformly continuous sub- and supersolutions of (\ref{lcp}) respectively, and
$u(x,T)\le v(x,T)$ for all $x$, then $u\le v$ everywhere.
\end{theorem}

It is clear that $\upen$ is a classical subsolution of (\ref{lcplog}),
\[
\min(-\mathcal{L} \upen, \upen-\psi) = 
\min({\mathsmaller{\frac{1}{\epsilon}}} \max(\psi-\upen,0), \upen-\psi) \le 0,
\]
and therefore also a viscosity subsolution, thus
$\upen\le u$ is a lower bound for the true solution.

\bigskip
We now seek to construct an upper bound by setting
\begin{eqnarray}
\widebar{u}^\epsilon &:=& \upen + \lambda^\epsilon, \\
\label{lagrmult}
\lambda^\epsilon &:=& \min\{\lambda\in\mathbb{R}, \lambda\ge 0: \, \upen + \lambda \ge \psi\} =  \max\{(\psi - \upen)^+\}.
\end{eqnarray}
Indeed, $\widebar{u}^\epsilon$ is a (classical and viscosity) supersolution of (\ref{lcplog}),
\[
\min(-\mathcal{L} \widebar{u}^\epsilon, \widebar{u}^\epsilon-\psi) = 
\min(r \lambda^\epsilon + \epsilon \max(\psi-\upen,0), 
\widebar{u}^\epsilon-\psi) \ge 0,
\]
and therefore $\widebar{u}^\epsilon\ge u$.
\bigskip

\begin{figure}[t]
\centering
\includegraphics[width=.45\columnwidth,height=6cm]{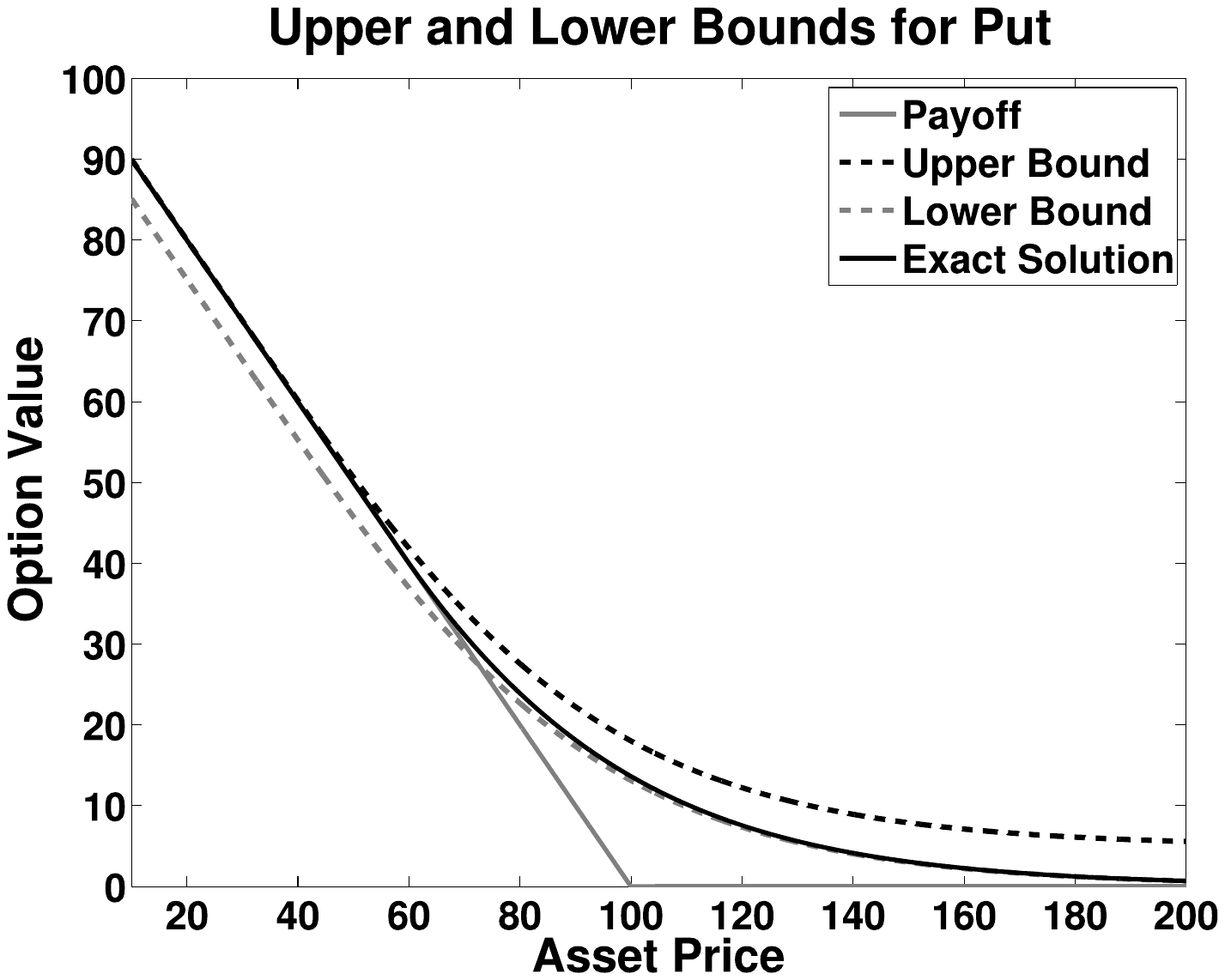}
\hfill
\includegraphics[width=.45\columnwidth,height=6cm]{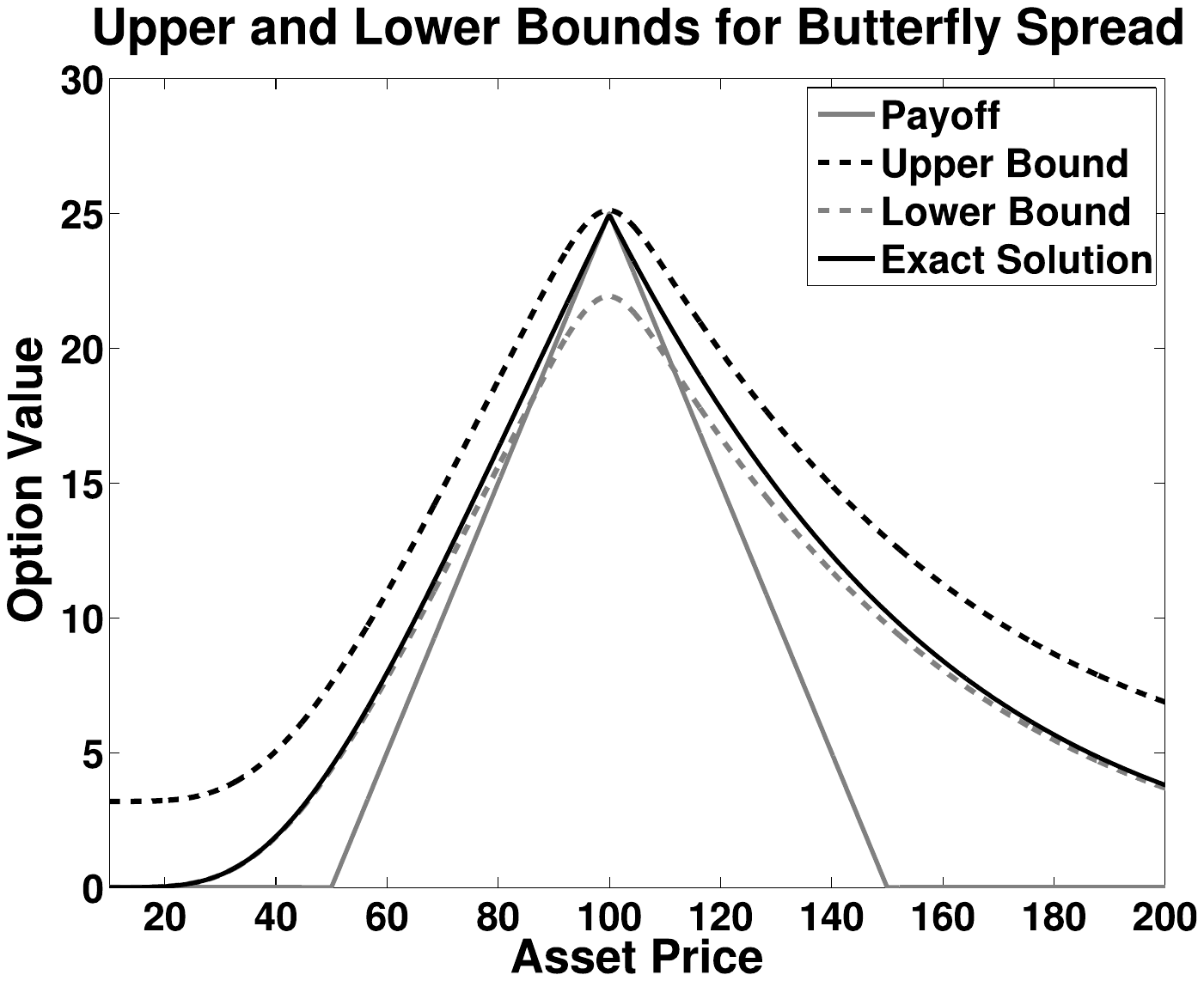}
\caption{Illustration of the lower bound $\Vpen$, upper bound $\overline{V}^\epsilon = \Vpen + \max\{(\Psi - \Vpen)^+\}$, payoff $\Psi$, and true value function $V$
for the American put (left) and the American butterfly spread (right), for $\epsilon = 100$ (put) and $\epsilon = 0.00005$ (butterfly). We are thinking of $\epsilon$ as a \emph{small} number, however for very small values, the bounds for the put become optically indistinguishable from the solution.}
\label{fig:UpperLowerBounds}
\end{figure}

From (\ref{maxbnd}) and the discussion thereafter,
we know that $\lambda^\epsilon$ can be estimated from (\ref{lagrmult})
by using the right-hand side from (\ref{maxbnd}) and values at concave kinks
and boundaries. We can use this fact to compute simple lower and upper bounds, which converge to the true solution.
Figure \ref{fig:UpperLowerBounds} illustrates this for the put and butterfly.
Note the different magnitude of the penalty parameter required for the put and butterfly to achieve similar accuracy.

\bigskip

The upper bounds are closely related to the regularised Lagrange multiplier approximation
in \cite{Ito_Kunisch_ParVarIneq_LagrangeMultiplier},
who propose to solve
\[ 
- \mathcal{L} u =  \max\big((\psi-u)/\epsilon+\bar{\lambda},0\big)
\]
for some fixed function $\bar{\lambda}>0$ large enough to make the solution feasible. 
This essentially corresponds to $\lambda^\epsilon = \epsilon \bar{\lambda}$
in (\ref{lagrmult}) and will thus be possible if the penalisation error is $O(\epsilon)$.

\bigskip

In the following section, we show that the order of $\lambda^\epsilon$ is either $O(\epsilon)$
in the case of no `active' concave kinks (i.e., where no non-convex kink lies in the active set of the inequality constraint)
or $O(\epsilon^{1/2})$ in the case of `active' concave kinks, as
expected from the asymptotic expansions in Section \ref{sec:asymp},
specifically \ref{subsec:putasymp} for the put (no concave kink) and \ref{subsec:butterfly} for the butterfly (active concave kink).

\subsection{Convergence Rates and Further Properties}

The following results follow directly from the comparison principle.
\begin{la}\label{Monotonicity_Lemma}
Denote by $u$ and $\widehat{u}$ the solutions to (\ref{lcplog}) with
obstacles $\psi$ and $\widehat{\psi}$, and, similarly, denote by $\upen$ and $\widehat{u}^{\epsilon}$ the corresponding solutions to ({lcplogpen}).
If $\psi\leq\widehat{\psi}$ everywhere, then we have
\begin{equation*}
\upen\leq\widehat{u}^{\epsilon}\quad\text{and}\quad u\leq \widehat{u}.
\end{equation*}
Trivially, $\upen$ and $u$ are nonnegative if $\psi\geq0$.
Moreover, denote by $u^{\epsilon_1}$ and $u^{\epsilon_2}$ the solutions to (\ref{lcplogpen})
corresponding to penalty parameters $\epsilon_1>\epsilon_2>0$, respectively. Then
$u^{\epsilon_1}\geq u^{\epsilon_2}$. 
\end{la}
\begin{proof}
For the first part, consider
\[
\min(-\mathcal{L} u, \phi-u) \le \min(-\mathcal{L} u, \widehat{\phi}-u)
\]
and
\[
-\mathcal{L} \upen = \frac{1}{\epsilon} \max(\psi- \upen,0) \leq \frac{1}{\epsilon} \max(\widehat{\psi}-\upen,0),
\]
such that $u$ and $\upen$ are subsolutions to their governing equations with $\psi$ replaced by $\widehat{\psi}$.
Similarly, for the second part, apply the same argument to
\[
-\mathcal{L} u^{\epsilon_1} = \frac{1}{\epsilon_1} \max(\psi-u^{\epsilon_1},0) \leq  \frac{1}{\epsilon_2} \max(\psi-u^{\epsilon_1},0).
\]
\end{proof}

We can apply the framework of \cite{jakobsen06}, pp.~4--8,
to estimate $\lambda^\epsilon$ in (\ref{lagrmult}).
\begin{theorem}
If $\psi$ is Lipschitz continuous and piecewise $C^1$ with linear growth and
\begin{enumerate}
\item
convex kinks, then
\[
0\le u-\upen \le C \epsilon;
\]
\item
concave kinks, then
\[
0\le u-\upen \le C \epsilon^{1/2}.
\]
\end{enumerate}
\end{theorem}
\begin{proof}
This follows precisely the steps in the proof of Theorem 2.1 in \cite{jakobsen06}.
Although the context there is that of non-linear PDEs, the results are sufficiently abstract to
accommodate PIDEs given a comparison principle as ascertained by Theorem \ref{theo:phamcomp}.
The main steps are based on smoothing the payoff with mollifiers, and bounding the approximation
error in the two cases.
\end{proof}

\section{Solution of a Variational Formulation}

From the previous section, we know $\max_x |u(x,t)-\upen(x,t)| = O(\epsilon)$ for payoffs with convex kinks.
Combined with the differentiability of $\upen$ with respect to $x$,
where the size of the derivative is independent of $\epsilon$,
\[
\pda{u}{x}(x,t) = \frac{u(x+\sqrteps)-u(x,t)}{\sqrteps} + O(\sqrteps) =
 \frac{\upen(x+\sqrteps)-\upen(x,t)}{\sqrteps} + O(\sqrteps)
\]
allows us to estimate the derivative up to $\sqrteps$ by a finite difference, which naturally `regularises' the differentiation. 
We know e.g.\ for the put from the asymptotic expansion that convergence will be better behaved everywhere except
in a small neighbourhood (of width $\sqrteps$) of the exercise boundary.
For non-convex kinks, convergence will be slower.\bigskip

This section develops estimates of the penalisation error for the derivative directly, via analysis in the $H^1$ norm.
We follow here the set-up of \cite{Zhang_AmericanJumpPaper}, who show convergence of penalisation in jump-diffusion models, but do
not derive convergence orders.

\subsection{Set-up}

We study problems (\ref{lcplog}) and (\ref{lcplogpen}), but on a localised domain
$\Ol:= \{x\in\mathbb{R} : |x|<l\}$ with boundary
$\partial\Ol:= \{x\in\mathbb{R} : |x|=l\}$.
It would be possible to work on $\mathbb{R}$, but this would require us to introduces weighted norms to be able to deal with functions
that do not decay (sufficiently fast) for large $x$ (such that their Sobolev norms are well defined), making the variational formulation
more cumbersome to write out.
Instead, on the finite domain, we can 
use the standard (separable Hilbert) spaces
$\Hl:= L^2(\Ol)$ and $\Vl:= \{u\in \Hl : \partial u/\partial x\in \Hl\} = H^1(\Ol)$
\cite{Adams_SobolevSpaces}.
For $\Sigma\in\{\Hl\,,\Vl\}$, define $L^2(0,T;\Sigma)$ as the (separable Hilbert) spaces of
measurable functions $u:[0,T]\rightarrow \Sigma$ satisfying
$u(\cdot,t)\in \Sigma$ for almost every $t\in[0,T]$ and for which
$\int^T_0|u(\cdot,t)|^2_{\Sigma}\ dt < \infty$,
equipped with their canonical inner products
(cf.\,\cite{Lions_ProbAuxLimitesNonHomogens_Vol1}).
The (Banach) space $L^{\infty}(0,T;\varSigma)$ is defined to
contain all measurable functions $u:[0,T]\rightarrow \Sigma$ satisfying
$u(\cdot,t)\in \Sigma$ for almost every $t\in[0,T]$ and for which
$|u|_{L^{\infty}(0,T;\varSigma)}:=\text{ess\,sup}_{t\in[0,T]}\,|u(\cdot,t)|_{\varSigma}<\infty$
(cf.\,\cite{Lions_ProbAuxLimitesNonHomogens_Vol1}).
\bigskip

Now, for $u\in \Hl$ and $x\in\Ol\,$, define the jump operator
\begin{equation*}
 (\Bopl u)(x) := \lambda\Big[ \int_{z,z+x\in\Ol} u(x+z)\ \nu(z) \drm z - u(x)\Big].
\end{equation*}
For $u$, $v\in \Vl$\,, define the bilinear forms
\begin{align*}
\al(u,v):=&\ \frac{\sigma^2}{2}\int_{\Ol}\frac{\partial u}{\partial x}\frac{\partial v}{\partial x}\drm x
+\int_{\Ol} ruv \drm x - \int_{\Ol}\Big(\mu-\frac{\sigma^2}{2}\Big)\frac{\partial u}{\partial x}v \drm x,
\end{align*}
where $\mu=r-q-\omega\lambda$ and
\begin{equation*}
\bl(u,v):= -\int_{\Ol}(\Bopl u)v\drm x.
\end{equation*}

We assume in the following that
$\psi:\mathbb{R}\to\mathbb{R}$ is continuous and there exists a constant $M>0$ such that
\begin{equation}
|\psi(x)|\leq Me^{M|x|},\quad x\in\mathbb{R}, \label{UstarEqV_Eq1}
\end{equation}
then set
\begin{equation*}
f(x):= \lambda\int_{z,z+x\notin\Ol}\psi(x+z)\ \nu(z) \drm z,\quad x\in\Ol\,
\end{equation*}
and assume that $f\in \Hl$.\bigskip

The following two lemmas are taken from \cite{Zhang_AmericansJumps_PhDThesis}, to where we also refer for the proof of the subsequent theorem.

\begin{la}\label{Bl_BoundedOperator}
We have $\Bopl\in\mathcal{L}(\Hl\,,\Hl)$, i.e. $\Bopl:\Hl\rightarrow \Hl$ is a bounded linear operator.
\end{la}
\begin{proof}
See \cite{Zhang_AmericansJumps_PhDThesis}.
\end{proof}

\begin{la}\label{WeaklyCoercive}
There exist constants $\vartheta$, $\xi>0$ such that
\begin{equation*}
\al(u,u) + \bl(u,u) \geq \vartheta |u|^2_{\Vl} - \xi|u|^2_{\Hl}\,,\quad u\in \Vl\,.
\end{equation*}
\end{la}
\begin{proof}
See \cite{Zhang_AmericansJumps_PhDThesis}.
\end{proof}

We are now in a position to formulate the variational inequality the solution of which
is the value function of the American option.

\begin{prob}\label{local_VarIneq}
Find a function $u\in L^2\big(0,T;\Vl\big)$, $\partial u/\partial t\in L^2\big(0,T;\Hl\big)$, such that
\begin{equation*}
u(\cdot,T)= \psi,\ u(x,\cdot)=\psi(x)\text{ for }x\in\partial\Ol\,,\ u\geq\psi\text{ a.e. on } \Ol\times[0,T]
\end{equation*}
and, a.e. on $[0,T]$, it is
\begin{equation}
-\Big(\frac{\partial u}{\partial t},v-u\Big) +\al(u,v-u)+\bl(u,v-u)-(f,v-u\brackl\geq 0\label{local_VarIneq_Eq1}
\end{equation}
for all $v\in \Vl$ with $v\geq\psi$.
\end{prob}

We emphasise that the payoff function $\psi$ is not confined to the interval $[-l,l]$ and that the function $f$ is given by a nonlocal integral.

\begin{theorem}
There exists a unique solution $u$ to Problem \ref{local_VarIneq} with
$u\in L^{\infty}(0,T;\Vl)$. 
\end{theorem}
\begin{proof}
See \cite{Zhang_AmericansJumps_PhDThesis} or \cite{Zhang_AmericanJumpPaper}.
\end{proof}

\subsection{Penalisation and Basic Properties}


\begin{prob}\label{local_VarIneq_penalised}
Let $\epsilon>0$ and define $\beta(\cdot):=-(\psi-\cdot)^+$.
Find a function $\upen\in L^2\big(0,T;\Vl\big)$, $\partial \upen/\partial t\in L^2\big(0,T;\Hl\big)$, such that
\begin{equation*}
\upen(\cdot,T) = \psi,\ \upen(x,\cdot)=\psi(x)\text{ for }x\in\partial\Ol
\end{equation*}
and, a.e. on $[0,T]$, we have
\begin{equation}
 -\Big(\frac{\partial \upen}{\partial t},v\Big) +\al(\upen\,,v)+\bl(\upen\,,v)-(f,v\brackl + \frac{1}{\epsilon}\big(\beta(\upen),v\big)= 0\label{local_VarIneq_penalised_Ee1}
 \end{equation}
for all $v\in \Vl$\,.
\end{prob}

\begin{theorem}\label{PenalisationSolConvergence}
There exists a unique solution $\upen$ to Problem \ref{local_VarIneq_penalised}.
Furthermore, there exists a constant $C>0$, $C$ independent of $\epsilon$, such that
\begin{equation}
|\upen|_{L^{\infty}(0,T;\Vl)}
+
\frac{1}{\sqrteps}|\beta(\upen)|_{L^2(0,T;\Hl)}
+
\Big|\frac{\partial \upen}{\partial t}\Big|_{L^2(0,T;\Hl)}\leq C.\label{PenalisationSolConvergence_Eq1}
\end{equation}
As $\epsilon\to 0$, we have that
$\upen\to u$ strongly and $\partial \upen/\partial t\to \partial u/\partial t$ weakly in $L^2(0,T;\Hl)$, where $u$ is the solution to Problem \ref{local_VarIneq}.
\end{theorem}
\begin{proof}
A result of this form comes up naturally when using penalisation to prove the existence of a solution to a variational inequality. In this particular case, it can be directly obtained by adapting the proof
given in \cite{Zhang_AmericansJumps_PhDThesis} for the non-localised problem. Alternatively, one can slightly extend a similar result given in
\cite{Lions_ApplVarIneqStochControl}.
\end{proof}

\begin{remark}
\label{monrem}
The results of Lemma \ref{Monotonicity_Lemma} still hold for the localised variational problem, but we omit the proof of this.
\end{remark}

\subsection{The American Put and Other Payoffs with Convex Kinks}

The following result is an extension of the one in \cite{Lions_ApplVarIneqStochControl} to jump diffusion,
and to accommodate kinks.
In the proofs, we work with weak coercivity instead of coercivity, and account for the loss of regularity at kinks explicitly.

\begin{theorem}\label{PenErrorAmericanPut}
Consider an American put option, i.e., let $\psi$ be given by
\begin{equation*}
\psi(x)=(K-e^x)^+,\quad x\in\mathbb{R},
\end{equation*}
and suppose $f\in \Hl$\,. There exists a constant $C>0$, $C$ independent of $\epsilon$, such that
\begin{equation*}
|u-\upen|_{L^2(0,T;\Vl)} + |u-\upen|_{L^{\infty}(0,T;\Hl)} \leq \sqrteps C.
\end{equation*}
\end{theorem}
\begin{proof}
Again, we extend a proof that was given in \cite{Lions_ApplVarIneqStochControl} for standard parabolic variational inequalities in $H^1_0$.
All constants $C_i$\,, with $i$ an integer, are taken to be independent of $\epsilon$ and $t$.
Plugging $-\beta(\upen)\in \Vl$ into \eqref{local_VarIneq_penalised_Ee1} gives
\begin{equation*}
\Big(\frac{\partial \upen}{\partial t},\beta(\upen)\Big)
- \al\big(\upen\,,\beta(\upen)\big)
- \bl\big(\upen\,,\beta(\upen)\big)
+\big(f,\beta(\upen)\big) - \frac{1}{\epsilon}\big(\beta(\upen),\beta(\upen)\big)=0,
\end{equation*}
which is equivalent to
\begin{align*}
&\hspace{-2 cm} \Big(\frac{\partial\beta}{\partial t}(\upen),\beta(\upen)\Big)
- \al\big(\beta(\upen),\beta(\upen)\big)
- \frac{1}{\epsilon}\big(\beta(\upen),\beta(\upen)\big)\\
=&\ \al\big(\psi,\beta(\upen)\big) + \bl\big(\upen\,,\beta(\upen)\big)-\big(f,\beta(\upen)\big)
-\Big(\frac{\partial\psi}{\partial t},\beta(\upen)\Big).
\end{align*}
Note that $\psi$ is independent of $t$. Integrating from $t$ to $T$, $t\in[0,T]$, we obtain
\begin{align}
& \hspace{-2 cm} \frac{1}{2}\big|\beta\big(\upen(t)\big)\big|^2_{\Hl} + \int^T_t \al\big(\beta(\upen),\beta(\upen)\big) \drm s
+ \frac{1}{\epsilon}|\beta(\upen)|^2_{L^2(t,T;\Hl)}\nonumber\\
=& -\int^T_t \al\big(\psi,\beta(\upen)\big) + \bl\big(\upen\,,\beta(\upen)\big)-\big(f,\beta(\upen)\big)
\drm s.\label{PennErr_PsiInH2_Eq1}
\end{align}
Recall Lemma \ref{Bl_BoundedOperator} and note that
\begin{eqnarray}
\nonumber
\hspace{-2 cm}
-\int^T_t \al\big(\psi,\beta(\upen)\big)\drm s 
&=& \int^T_t\Big(\,
\frac{\sigma^2}{2}\int_{\Ol} \frac{\partial\psi}{\partial x}\,\frac{\partial}{\partial x}(\psi-\upen)^+\drm x\\
&+& r\int_{\Ol}\psi\,(\psi-\upen)^+\drm x
- (\mu-\frac{\sigma^2}{2})\int_{\Ol}\frac{\partial \psi}{\partial x}\,(\psi-\upen)^+ \drm x
\Big) \drm s,
\label{anothereqn}
\end{eqnarray}
in which
\begin{align*}
& \int^T_t\Big(\, \int_{\Ol} \frac{\partial\psi}{\partial x}\,\frac{\partial}{\partial x}(\psi-\upen)^+\drm x\Big)\drm s
= -\int^T_t\Big(\int_{\Ol} \frac{\partial^2\psi}{\partial x^2}\,(\psi-\upen)^+\drm x\Big)\drm s.
\end{align*}
As $\psi$ is known explicitly, we can write
\begin{align}
& \hspace{-1 cm} \int^T_t \,
\int_{\Ol} \frac{\partial\psi}{\partial x}\,\frac{\partial}{\partial x}(\psi-\upen)^+\drm x \drm s\label{PenErrorAmericanPut_Eq0.95}\\
=&\,
\int^T_t\Big(
-\Big[e^x(\psi-\upen)^+\Big]^{\log K}_{-l}+
\int^{\log K}_{-l} e^x\,(\psi-\upen)^+\drm x
 \Big) \drm s\label{PenErrorAmericanPut_Eq1}\\
\leq&\,
\int^T_t\Big(\,
\int^{\log K}_{-l} e^x\,(\psi-\upen)^+\drm x
\Big) \drm s
\; \leq \;
K\int^T_t\Big(\,
\int_{\Ol} (\psi-\upen)^+\drm x
\Big) \drm s.\label{PenErrorAmericanPut_Eqq2}
\end{align}
To get from \eqref{PenErrorAmericanPut_Eq1} to \eqref{PenErrorAmericanPut_Eqq2}, we used the following fact: since $\upen\in L^2(0,T;\Vl)$, a monotonicity result in Remark \ref{monrem} gives that $\upen(s)\geq 0$ for almost every $s\in[0,T]$; hence, $\big(\psi(\log K)-\upen(\log K)\big)^+=\big(-\upen(\log K)\big)^+=0$ almost everywhere on $[0,T]$.
Having observed this, applying \eqref{PenErrorAmericanPut_Eqq2} to \eqref{anothereqn}, we then obtain
\begin{align}
-\int^T_t \al\big(\psi,\beta(\upen)\big) \drm s
\leq
C_0|\psi|_{H^1(\Ol)}\,|\beta(\upen)|_{L^2(t,T;\Hl)},\label{PenErrorAmericanPut_Eq2.5}
\end{align}
which, applied to \eqref{PennErr_PsiInH2_Eq1}, gives
 \begin{multline*}
 \int^T_t \al\big(\beta(\upen),\beta(\upen)\big)\drm s
 + \frac{1}{\epsilon}|\beta(\upen)|^2_{L^2(t,T;\Hl)}
 \leq C_1\Big( |\psi|_{H^2(\Ol)}\,|\beta(\upen)|_{L^2(t,T;\Hl)}\\
 + |\upen|_{L^2(t,T;\Hl)}\,|\beta(\upen)|_{L^2(t,T;\Hl)}
 + |f|_{\Hl}\,|\beta(\upen)|_{L^2(t,T;\Hl)}\Big).
 \end{multline*}
The splitting of the integral was necessary because of the kink of $\psi$,
whereas for $\psi\big|_{\Ol}\in H^2(\Ol)$ the last inequality follows directly
by integration by parts.
Applying Lemma \ref{WeaklyCoercive} and \eqref{PenalisationSolConvergence_Eq1} to the last expression, we then get
\begin{equation}
\frac{1}{\epsilon}|\beta(\upen)|_{L^2(t,T;\Hl)}\leq C_2\label{PennErr_PsiInH2_Eq2}
\end{equation}
for $0<\epsilon<1$. Next, applying Lemma \ref{WeaklyCoercive}, \eqref{PenalisationSolConvergence_Eq1} and \eqref{PennErr_PsiInH2_Eq2} to equation \eqref{PennErr_PsiInH2_Eq1} yields
\begin{equation}
|\beta(\upen)|_{L^2(t,T;\Vl)}
+ |\beta(\upen)|_{L^{\infty}(t,T;\Hl)}
\leq \sqrteps C_3\,.\label{PennErr_PsiInH2_Eq3}
\end{equation}
We define $\rpen:=\psi-u + (\psi-\upen)^-$, where $(\psi-\upen)^-:=-\min\{\psi-\upen,0\}$; in particular, this means $\upen-u = \rpen + \beta(\upen)$. Owing to \eqref{PennErr_PsiInH2_Eq3}, to prove the theorem, it is now sufficient to show
that
\[
|\rpen|_{L^2(t,T;\Vl)} + |\rpen|_{L^{\infty}(t,T;\Hl)} \leq \sqrteps C_4.
\]
We set $v= \rpen+u=\psi + (\psi-\upen)^- \geq \psi$ in \eqref{local_VarIneq_Eq1}
and $v= -\rpen\in \Vl$ in \eqref{local_VarIneq_penalised_Ee1} and sum the two expressions to obtain
\begin{equation*}
-\Big(\frac{\partial}{\partial t}(u - \upen),\rpen\Big) + \al(u - \upen\,,\rpen)
+ \bl(u - \upen\,,\rpen) + \frac{1}{\epsilon}\big(\beta(\upen),u-\psi\big) \geq 0.
\end{equation*}
As
$-\frac{1}{\epsilon}\big(\beta(\upen),u-\psi\big)\geq 0$,
we further get
\begin{align*}
-\Big(\frac{\partial}{\partial t}(\upen-u),\rpen\Big) + \al(\upen-u,\rpen)
+ \bl(\upen-u,\rpen) \leq 0,
\end{align*}
which in return gives
\begin{align*}
-\Big(\frac{\partial \rpen}{\partial t},\rpen\Big) + \al(\rpen\,,\rpen)
+ \bl(\rpen\,,\rpen) \leq \Big(\frac{\partial\beta}{\partial t}(\upen),\rpen\Big)
- \al\big(\beta(\upen),\rpen\big) - \bl\big(\beta(\upen),\rpen\big).
\end{align*}
We define $\widehat{\rpen}:=e^{\xi t} \rpen$\,, where we use $\xi$ from Lemma \ref{WeaklyCoercive}.
Multiplying both sides of the last inequality by $e^{2\xi t}$, we get
\begin{align*}
&\hspace{-1.5 cm} -\Big(\frac{\partial \widehat{\rpen}}{\partial t},\widehat{\rpen}\Big)
+ \xi(\widehat{\rpen}\,,\widehat{\rpen})
+ \al(\widehat{\rpen}\,,\widehat{\rpen})
+ \bl(\widehat{\rpen}\,,\widehat{\rpen})\\
\leq&\ \Big(\frac{\partial}{\partial t}\big[e^{\xi t}\beta(\upen)\big],\widehat{\rpen}\Big)
-  \xi\big(e^{\xi t}\beta(\upen),\widehat{\rpen}\big)
- \al\big(e^{\xi t}\beta(\upen),\widehat{\rpen}\big)
- \bl\big(e^{\xi t}\beta(\upen),\widehat{\rpen}\big).
\end{align*}
Noting that $\hat{\rpen}(T) = e^{\xi T}\big[ \psi(T)-u(T) +\big(\psi(T)-\upen(T)\big)^-\big] = 0$,
integrating from $t$ to $T$, $t\in[0,T]$, gives
\begin{multline}
\frac{1}{2}|\widehat{\rpen}(t)|^2_{\Hl}
+ \int^T_t \xi(\widehat{\rpen}\,,\widehat{\rpen}) + \al(\widehat{\rpen}\,,\widehat{\rpen})
+ \bl(\widehat{\rpen}\,,\widehat{\rpen})\drm s\\
\leq
-\big(e^{\xi t}\beta(\upen)(t),\widehat{\rpen}(t)\big)
- \int^T_t e^{\xi t}\Big(\beta(\upen),\frac{\partial\widehat{\rpen}}{\partial t}\Big)\drm s\\
 - \int^T_t \xi\big(e^{\xi t}\beta(\upen),\widehat{\rpen}\big)+
\al\big(e^{\xi t}\beta(\upen),\widehat{\rpen}\big)
+ \bl\big(e^{\xi t}\beta(\upen),\widehat{\rpen}\big) \drm s.\label{PennErr_PsiInH2_Eq4}
\end{multline}
We now make three observations, which, taken together, will yield the desired result.
First, according to Lemma \ref{WeaklyCoercive}, we have
\begin{equation*}
\int^T_t \xi(\widehat{\rpen}\,,\widehat{\rpen}) + \al(\widehat{\rpen}\,,\widehat{\rpen})
+ \bl(\widehat{\rpen}\,,\widehat{\rpen})\drm s \geq \vartheta\int^T_t |\widehat{\rpen}|^2_{\Vl}\drm s.
\end{equation*}
Second, according to \eqref{PenalisationSolConvergence_Eq1},
\begin{align*}
&\hspace{-1.5 cm} -\int^T_t \Big(e^{\xi t}\beta(\upen),\frac{\partial\widehat{\rpen}}{\partial t}\Big)\drm s\\
\leq&\
\xi e^{2\xi T}|\beta(\upen)|_{L^2(t,T;\Hl)}\,|\rpen|_{L^2(t,T;\Hl)} + e^{2\xi T}|\beta(\upen)|_{L^2(t,T;\Hl)}\,\Big|\frac{\partial u}{\partial t}\Big|_{L^2(t,T;\Hl)}\\
\leq&\ 
C_5\,|\beta(\upen)|_{L^2(t,T;\Hl)}\,|\rpen|_{L^2(t,T;\Hl)} + \epsilon C_6\,.
\end{align*}
Third, we have
\begin{align*}
&\hspace{-1.5 cm}-\big(e^{\xi t}\beta(\upen)(t),\widehat{\rpen}(t)\big)
- \int^T_t \xi\big(e^{\xi t}\beta(\upen),\widehat{\rpen}\big) + a\big(e^{\xi t}\beta(\upen),\widehat{\rpen}\big)
+ b\big(e^{\xi t}\beta(\upen),\widehat{\rpen}\big)\drm s\\
\leq&\ 
C_7 |\beta(\upen)(t)|_{\Hl}\,|\rpen(t)|_{\Hl}
+ C_8 |\beta(\upen)|_{L^2(t,T;\Vl)}\,|\rpen|_{L^2(t,T;\Vl)}.
\end{align*}
Applying the last three statements as well as \eqref{PennErr_PsiInH2_Eq3} to \eqref{PennErr_PsiInH2_Eq4} completes the proof.
\end{proof}

Finally, we formulate a corollary which states that the result just given for the American put also holds for a wider class of functions including a number of traditional option payoffs.

\begin{cor}\label{PenaltyError_KinkedPayoffs}
If there is a finite number of
disjoint open intervals $I_i:=(x_i\,,x_{i+1})$, $0\leq i\leq N$, such that $\bigcup^N_{i=0}[x_i\,,x_{i+1}]=[-l,l]$, $\psi\big|_{I_i}\in H^2(I_i)$ for $0\leq i\leq N$, and, additionally,
\begin{equation*}
\lim_{x\uparrow x_i} \frac{\partial \psi}{\partial x}(x_i)\leq \lim_{x\downarrow x_i} \frac{\partial \psi}{\partial x}(x_i)
\end{equation*}
for $1\leq i\leq N$, then the result of
Theorem \ref{PenErrorAmericanPut} also holds. In particular, this includes piecewise smooth functions which are convex,
e.g., a straddle and an American call.
\end{cor}
\begin{proof}
Integrating by parts, we can write \eqref{PenErrorAmericanPut_Eq0.95} as
\begin{equation*}
\int_{\Ol} \frac{\partial\psi}{\partial x}\,\frac{\partial}{\partial x}(\psi-\upen)^+\drm x
= \sum_{i=0}^N\Big[\frac{\partial\psi}{\partial x}(\psi-\upen)^+\Big]^{x_{i+1}}_{x_i}
-\sum_{i=0}^N\int_{I_i}\frac{\partial^2 \psi}{\partial x^2}(\psi-\upen)^+\drm x,
\end{equation*}
in which the first term on the right hand side equals
\begin{equation*}
\sum_{i=1}^N
\Big(
\lim_{x\uparrow x_i}\frac{\partial\psi}{\partial x}(x)\big(\psi(x_i)-\upen(x_i)\big)^+
-
\lim_{x\downarrow x_i}\frac{\partial\psi}{\partial x}(x)\big(\psi(x_i)-\upen(x_i)\big)^+
\Big)\leq 0,
\end{equation*}
and we can replace \eqref{PenErrorAmericanPut_Eq2.5} by
\begin{equation*}
-\int^T_t \al\big(\psi,\beta(\upen)\big) \drm s
\leq
C_{7}
\sum_{i=0}^N |\psi|_{H^2(I_i)}
|\beta(\upen)|_{L^2(t,T;\Hl)}.
\end{equation*}
Having done this, we then proceed as in the proof of Theorem \ref{PenErrorAmericanPut}.
\end{proof}

\begin{remark}
We have shown convergence of order 1/2 in $\epsilon$ in the $L^2(0,T;H^1)$ norm for
piecewise smooth obstacles with convex kinks, while only convergence (but no positive convergence order) can be shown
for non-convex kinks.
Because of the embedding of $H^1$ in $L^\infty$ in one dimension, this implies the same convergence orders in the
maximum norm, which are weaker results than the higher orders (1 and 1/2, respectively) established in Section \ref{sec:comparison}
via viscosity techniques.

\bigskip

The above result further implies convergence of order 1/2 of the derivative in the $L^2(0,T;L^2)$ norm, which is a new result and does not follow from the one in Section \ref{sec:comparison}.
Specifically,
\begin{eqnarray*}
C^2 \epsilon &\ge&
|u-u^\epsilon|_{L^2(0,T;H^1(\Omega))}^2 \\
&\ge& 
\int_0^T \int_{\Omega} \left( \frac{\prt u}{\prt x}- \frac{\prt u^\epsilon}{\prt x}  \right)^2 \, {\rm d} x \, {\rm d} t \\
&=& \int_0^T \int_{\Omega_S} S \left( \frac{\prt V}{\prt S}- \frac{\prt V^\epsilon}{\prt S}  \right)^2 \, {\rm d} S \, {\rm d} t,
\end{eqnarray*}
where $\Omega_S$ is the image of $\Omega$ under transformation into $S$ coordinates, $S=S_0 \exp(x)$.
Comparing this to (\ref{semi-norm}) we see that the variance of the hedging error will behave like $O(\epsilon)$.
\end{remark}

\section{Discussion and Applications}

\bigskip

\subsection{Interplay Between Penalisation and Discretisation}

A comment is due on the effect of discretisation of the underlying PDE on the penalisation error, and, conversely,
of penalisation on the discretisation error.

\subsubsection*{Penalisation of discrete systems}

Here, we reconcile the fact that convergence of penalised solutions to finite-dimensional (discretised) variational inequalities is 
almost always of first order in the penalty parameter irrespective of the payoff (see, e.g., \cite{ForsythQuadraticConvergence}),
with the observation of the earlier sections of a clear difference between different payoff classes in both theory and numerical results.

\bigskip

We consider the Black-Scholes case and a discretisation with
equally spaced mesh points $S_i=i h$, $1\le i\le N$,
with mesh width $h$.
The standard central difference scheme with fully implicit timestepping with time step $k$ can be written as 
\begin{eqnarray}
\label{bslcp} \\ \nonumber
\min\left( -\frac{V^{j+1}_i-V_i^j}{k} -
\frac{1}{2} \sigma^2 S_i^2 \frac{V^j_{i+1}-2 V^j_i + V^j_{i-1}}{h^2} - r S_i \frac{V^j_{i+1}- V^j_{i-1}}{2h} + k r V^j_i, \,
V_i^j - \Psi(S_j)  \right) = 0,
\end{eqnarray}
where $\Psi$ is the payoff function and $V_i^j$ is the finite difference approximation to $V$ at mesh point $i h$ and time $j k$.
Stepping backwards in time, in each time step, one has to solve a discrete linear complementarity problem of the form
\begin{equation}
\label{eqn:discrlcp}
\min(A x-b,x-c) = 0,
\end{equation}
where
\begin{equation}
\label{findiffscheme}
(A x)_i = x_i - k \frac{1}{2} \sigma^2 S_i^2 \frac{x_{i+1}-2 x_i + x_{i-1}}{h^2} - k r S_i \frac{x_{i+1}- x_{i-1}}{2h} + k r x_i,
\end{equation}
such that $A\in \mathbb R^{N\times N}$ is typically an M-matrix (subject to conditions on $\sigma$ and $r$, and can be forced to be an M-matrix by
selective upwinding, see \cite{Wang08maximaluse}),
$b, c, x\in \mathbb R^{N}$ (we assume $x_0$ and $x_{N+1}$ are fixed by boundary conditions).
\bigskip

Note that to obtain (\ref{eqn:discrlcp}) we have multiplied the first term in (\ref{bslcp}) by $k$, which was allowable
because the solution of (\ref{eqn:discrlcp}) for fixed $N$ is invariant to scaling of the two arguments of the `$\min$'
by a positive constant.
However, scaling does become relevant for picking an appropriate penalty parameter for the disretised system
(see also \cite{huangetal11} for inexact arithmetic considerations surrounding this issue).
In particular, \cite{ForsythQuadraticConvergence} consider a penalised equation
\begin{eqnarray}
\label{pendiscr}
A \xpen-b = \lar \max(c-\xpen,0),
\end{eqnarray}
for a large positive parameter $\lar$,
and show that
\begin{equation}
\label{pendiscrest}
\|x-\xpen\| \le C/\lar,
\end{equation}
where $\|\cdot \|$ is the maximum norm.
The error bound in (\ref{pendiscrest}) is of first order in the penalty parameter $1/\lar$.
\bigskip

We now explain why this does not contradict the discrepancy between convex and concave payoffs found in the previous sections.
A key estimate on p.~2117 in \cite{ForsythQuadraticConvergence} is
\[
\|A c\| \le const,
\]
for some positive constant, where $c=(\Psi(S_i))_i$ is the discretised payoff.
Applied to a Lipschitz payoff $\Psi$, the first central difference from (\ref{findiffscheme}) is bounded as $h\rightarrow 0$,
and the second finite differences is $O(1/h)$ (with its maximum in the vicinity of kinks).
So as $A$ contains these spatial finite differences \emph{multiplied} by $k$, 
$\|A c\|_\infty = \max_i |(A c)_i| = O(k/h)$, and therefore, as long as $k/h$ is kept fixed,
$C$ in (\ref{pendiscrest}) is independent of the mesh size.
(Note that keeping $k$ proportional to $k$ is a sensible refinement regime as the Crank-Nicolson
central difference scheme has consistency order 2 in both $k$ and $h$ and is unconditionally stable.)
\bigskip

We now elucidate the relation between $\lar$ and $\epsilon$.
For $k,h\rightarrow 0$,
the above penalised equation is \emph{not} consistent (in the classical sense of consistency of finite difference schemes)
with the penalised PDE (\ref{pen}) with fixed penalty parameter $\lar$. Instead,
if we arrange (\ref{pendiscr}) into
\begin{eqnarray*}
\frac{1}{k} (A \xpen)_i - \frac{1}{k}b_i &=& 
\frac{x_i-b_i}{k} - \frac{1}{2} \sigma^2 S_i^2 \frac{x_{i+1}-2 x_i + x_{i-1}}{h^2} -  r S_i \frac{x_{i+1}- x_{i-1}}{2h} + r x_i \\
&=&
\frac{\lar}{k} \max(c_i-\xpen_i,0),
\end{eqnarray*}
the `effective' penalty parameter
is $\lar/k$ and increases with $k\rightarrow 0$, so
substituting back $x_i=V^j_i$, $b_i=V^j_{i+1}$ $c_i=\Psi(S_i)$, one sees that
(\ref{pendiscr})
is consistent with the obstacle problem (\ref{lcp}) itself.
If we replace $\lar$ by $k/\epsilon$ in (\ref{pendiscr}) to get a scheme consistent with the penalised PDE (\ref{pen}),
(\ref{pendiscrest}) still holds, however, the constant $C$ generally
depends on $k$ and $h$.
Retracing the steps leading up to the key bound for the penalisation error, (A.6) in \cite{ForsythQuadraticConvergence}, one finds that only the positive part $\max(A c,0)$ is relevant for the estimate and not $\|Ac\|$,
so for the (convex) put payoff, in particular, $C$ in (\ref{pendiscrest}) is still asymptotically independent of the mesh size.
For the butterfly (with a concave kink), in contrast, $\|\max(A c,0)\|=O(1/h)$ and $C$ goes to $\infty$ for $h\rightarrow 0$, $k/h$ fixed.
This reflects the fact that, for the butterfly, the limiting continuous problem exhibits reduced convergence order in $\epsilon$.
\bigskip

Therefore, the analysis of the limiting continuous problem informs the choice of penalty parameter for the discretised system.
This is very clearly seen from Figure \ref{fig:UpperLowerBounds}.

\subsubsection*{Smoothing and discretisation of penalised equations}

We now turn the order of discretisation and penalisation around and consider the discretisation of a penalised PDE.
The penalised PDE (\ref{pen}) does not have a (known) closed-form solution and has to be solved numerically.
Error estimates for a finite element  approximation to the penalised heat equation have been given, e.g., by
\cite{scholz86} and \cite{witte09},
\begin{eqnarray}
\label{fempen}
\|\upen - \widehat{u}^{\epsilon} \| \le \left(c + \frac{C}{\sqrteps}\right) \cdot \left(k+h^2\right),
\end{eqnarray}
where $h$ is the mesh size and $k$ the timestep of an implicit Euler or $\theta$-method respectively, and $\|\cdot\|$ the $L_2$ norm.
\bigskip

In contrast, error bounds for the unpenalised problem found in
\cite{allegretto01} have reduced order in $k$ and $h$,
\begin{eqnarray}
\label{femunpen}
\|u - \widehat{u}\| \le c  \cdot \left(k^{1/2} +h\right).
\end{eqnarray}

\bigskip
These results reflect the fact that penalisation smooths the solution.
Consequently, the finite element error bounds (\ref{fempen}) deteriorate for decreasing
$\epsilon$, and the order of convergence in the mesh parameters is lower for 
the limiting variational inequality. 
The above results are based on the assumption of sufficiently smooth obstacles, such that the penalisation error is determined by smooth pasting at the free boundary and not
any kinks of the payoff.
\bigskip

This technique of smoothing the solutions to non-linear PDEs by penalisation in order to derive grid convergence rates for the limiting problem is used in the more general context of HJB and Isaacs equations in \cite{jakobsen06}.
\bigskip

We should remark that, in \emph{practice}, one can observe $O(k^{3/2}+h^2)$ convergence for Crank-Nicolson time stepping ($\theta=1/2$) even in the limit ($\epsilon\rightarrow 0$), i.e., for the direct discretisation
of (\ref{lcp}).
The convergence is even $O(k^2+h^2)$ for a suitably adapted time stepping scheme, see \cite{ForsythQuadraticConvergence,ReisingerWhitley}, which accounts for the singular behaviour of the solution close to expiry.
So in practice, one can let $\epsilon \rightarrow 0$ without negatively affecting mesh convergence (subject to machine precision effects, see \cite{huangetal11}).

\bigskip

\subsection{Richardson Extrapolation}
\label{subsec:rich}

We now show how extrapolation using the asymptotic results can be used to 
generate more accurate numerical solutions.

\bigskip
Consider here the American put.
We know from Section \ref{subsec:putasymp} (see also Table \ref{Tab:PenErrRatesSumm}) that the leading order correction to the penalty solution is proportional to $\epsilon$. So
doing the calculation with $\epsilon$ and  $2 \epsilon$, then taking one twice minus
the other,
\[
\widehat{V}^{\epsilon} = 2 \Vpen - V^{2\epsilon},
\]
will be a second order approximation to $V$ (assuming the next term in the expansion is quadratic).

\bigskip
For the finite difference computation of $\Vpen$, we choose a mesh size $h \sim \sqrteps$, for two reasons.
The (empirically observed) finite difference error is $O(h^2)$, whereas the penalisation error is $O(\epsilon)$, so the above choice makes both terms
the same order of magnitude.
Secondly, although the convergence of the penalised PDE solution is $O(\epsilon)$, overlaid is a displacement of the exercise boundary by $\sqrteps$
(see Section \ref{subsec:putasymp}), at which the penalisation error changes rapidly,
so extrapolation of the continuous equation (or, in practice, one with very small fixed grid size) does \emph{not} result in an order improvement
of the maximum error.
However, extrapolation
with mesh width $O(\sqrteps)$ coupled to the penalisation, gives the desired numerical results. This is because the `inner region' is not resolved
within grid cells of width $O(\sqrteps)$ and therefore does not destroy the convergence order of Richardson extrapolation.
The results are summarised in Table \ref{Tab:PenErrRatesExtra}. Note that by this procedure we gain a full convergence order in the derivative as well.

\begin{table}[t]
\begin{tabular}{|c||c|c|c|c||c|c|c|c|}
\multicolumn{1}{c}{}
& \multicolumn{1}{c}{Value} &\multicolumn{1}{c}{Delta}  \\
\hline
Penalty & 1.0000 & 0.4815  \\
Extrapolation & 2.0061 & 1.5015 \\
\hline
\end{tabular}
\caption{Order of convergence with respect to the penalty parameter, in the maximum norm, for original and extrapolated value and its derivative.
The setting is the Black-Scholes model with parameters as earlier.
}
\label{Tab:PenErrRatesExtra}
\end{table}


\bigskip
The inner (asymptotic) analysis in Section \ref{subsec:putasymp} is independent of the
volatility to the order of accuracy we have given. One could use a local volatility model and simply freeze the volatility
at its local value. 
So even in non-Black--Scholes models,
Richardson extrapolation may be a good way of using this to get a more accurate
outer put value with little extra effort.

The strategy should also work for multi-factor models.

\subsection{Extensions}

While the analysis in this article focuses on Black-Scholes and jump-diffusion models, the main results, especially of Section \ref{sec:comparison} and the applicability of extrapolation, 
should extend to other settings, including local volatility models and derivatives on more than one underlying or on an asset modelled by additional stochastic factors, e.g., stochastic volatility or interest rates.
Another interesting extension would be to free-boundary problems arising from portfolio selection under transaction costs, but we anticipate especially the matched asymptotic expansions to differ more substantially here due to the presence of first order derivatives in the penalty term (c.f.\ \cite{normandavis}).

\bibliography{penalty}
\bibliographystyle{plain}

\end{document}